\documentclass[fleqn,usenatbib,useAMS]{mnras}

\usepackage[T1]{fontenc}
\usepackage{ae,aecompl}

\usepackage{graphicx}	
\usepackage{amsmath}	
\usepackage{amssymb}	

\usepackage{xcolor}
\usepackage{footnote}
\usepackage{footmisc}
\usepackage[para]{threeparttable}
\usepackage{natbib}
\usepackage{amssymb}
\usepackage{amsmath}
\usepackage{longtable}
\usepackage[innercaption]{sidecap}
\usepackage{etoolbox}
\robustify\tnote
\usepackage{rotating}
\newcommand{\degree}{\ensuremath{^\circ}}

\title[Dense Gas towards Galactic TeV Sources]{Dense Molecular Gas at 12\,mm Towards Galactic TeV gamma-ray Sources}

\author[P. de Wilt et al.]{
P. de Wilt,$^{1}$\thanks{E-mail: phoebe.dewilt@gmail.com}
G. Rowell,$^{1}$
A.J. Walsh,$^{2}$
M. Burton,$^{3,4}$
J. Rathborne,$^{5}$
Y. Fukui,$^{6}$\newauthor
A. Kawamura$^{7}$
and F. Aharonian$^{8,9,10}$
\\
$^{1}$School of Physical Sciences, University of Adelaide, Adelaide, 5005, Australia\\
$^{2}$International Centre for Radio Astronomy Research, Curtin University, Bentley, 6102, Australia\\
$^{3}$School of Physics, University of New South Wales, Sydney, 2052, Australia\\
$^{4}$Armagh Observatory and Planetarium, College Hill, Armagh BT61 9DG, Northern Ireland\\
$^{5}$CSIRO Astronomy and Space Science, Marsfield, 2122, Australia\\
$^{6}$Department of Astrophysics, Nagoya University, Furocho, Chikusa-ku, Nagoya, Aichi, 464-8602, Japan\\
$^{7}$National Astronomical Observatory of Japan, 2-21-1 Osawa, Mitaka, Tokyo 181-8588 Japan\\
$^{8}$Dublin Institute for Advanced Studies, School of Cosmic Physics, 31 Fitzwilliam Place, Dublin 2, Ireland\\
$^{9}$Max Planck Institute for Nuclear Physics, Saupfercheckweg 1, 69117 Heidelberg, Germany\\
$^{10}$National Research Nuclear University (MEPHI), 115409, Moscow, Russia
}

\date{Accepted XXX. Received YYY; in original form ZZZ}

\pubyear{2016}

\begin{document}
\label{firstpage}
\pagerange{\pageref{firstpage}--\pageref{lastpage}}
\maketitle

\begin{abstract}
The High Energy Stereoscopic System (H.E.S.S.) has revealed many TeV ($10^{12}$eV) gamma-ray sources along the Galactic Plane and around 30 $\%$ of these sources remain unidentified. The morphology and dynamics of dense gas coincident and surrounding the gamma-ray emission can provide clues about the nature of the TeV emission. The H$_2$O Southern Galactic Plane Survey (HOPS) undertaken with the Mopra radio telescope, includes several dense gas tracers, such as NH$_3$\,(n,n) transitions and HC$_3$N\,(3-2), star formation tracers including H$_2$O masers, and radio recombination lines which trace ionised gas. A search for dense gas, traced by NH$_3$\,(1,1) emission seen in HOPS and additional observations, towards Galactic TeV sources has been undertaken. Of the 49 Galactic TeV sources covered by 12\,mm observations, NH$_3$\,(1,1) is detected towards or adjacent to 38 of them. Dense gas counterparts have been detected near several unidentified Galactic TeV sources which display morphology pointing to a hadronic origin to the TeV gamma-rays. The dense gas detected towards some TeV sources displays unusual emission characteristics, including very broad linewidths and enhanced ortho-to-para NH$_3$ abundance ratios towards HESS\,J1745$-$303 and HESS\,J1801$-$233, which reflects previous shock activity within the gas.
\end{abstract}




\section{Introduction}
\label{introduction}

The sensitivity of the current generation of ground based VHE gamma-ray experiments such as the High Energy Spectroscopic System (H.E.S.S., \citealt{2009AIPC.1112...16D}), has led to the discovery of many previously unknown sources of TeV ($10^{12}$eV) gamma-rays towards the Galactic plane. The largest fraction of Galactic TeV gamma-ray sources is still made up of unidentified objects \citep{2015arXiv150903872P}.

TeV gamma-rays are produced by interactions of relativistic particles and so their observation provides a probe of non-thermal astrophysical processes. GeV-TeV gamma-rays are produced hadronically, through proton-proton interactions, and/or leptonically, via Inverse Compton scattering or Bremsstrahlung. The presence of dense molecular gas can give clues about both leptonic and hadronic production of TeV gamma-rays. 

The flux of gamma-rays which are produced hadronically, through proton-proton interactions, can be expected to peak along with the density of gas near proton accelerators. Coincident dense gas and TeV emission can be seen in sources such as HESS\,J1801$-$233 \citep[see][]{2008A&A...481..401A}, a supernova remnant (SNR) / molecular cloud interaction region, to the NE of SNR W28, as well as parts of HESS\,J1745-303 \citep{2008A&A...483..509A} and RX J1713.7-3946 \citep{2012MNRAS.422.2230M}. TeV gamma-ray sources which are produced leptonically through inverse Compton emission, for example pulsar wind nebulae (PWNe), often exhibit asymmetric morphology. In the case of PWNe, this asymmetry may be explained by an inhomogeneous distribution of interstellar gas around pulsars which strongly influences their development \citep[e.g.][]{2001ApJ...563..806B}. The TeV emission then tails off in a direction away from the denser parts of the interstellar gas. 

Star formation that may not show up as infra-red sources may be traced by molecular transitions in the 12\,mm band pass, such as H$_2$O masers \citep{2011MNRAS.416.1764W}. There is evidence that H$_2$O masers, signposts of outflows and shocked gas, may be observable at very early stages of star formation \citep[e.g][]{2000ApJ...530..371F}. Ionised gas in H{\sc ii} regions is also traced, by radio recombination lines such as H69$\alpha$ which is included in our study. This provides an opportunity to test what role star forming regions play in TeV gamma-ray emission.

High mass stars have been linked with TeV gamma-ray emission through particle acceleration in SNRs and PWNe. High mass star forming regions and/or young high mass stellar objects may also have the potential to produce observable TeV gamma-ray emission through several mechanisms. The regions themselves could be sites of particle acceleration or the dense gas associated with these regions could simply provide target material for accelerate particles. Protostellar jets (which could be traced by H$_2$O masers), have been identified as potential sources of observable GeV gamma-rays \citep[see][]{2012AIPC.1505..281A} where particles are accelerated at the jet termination shock \citep[e.g.][]{2010A&A...511A...8B}. Colliding wind binaries have been linked to TeV emission \citep[e.g.][]{2006MNRAS.372..801P,2011A&A...530A..49B}, but, so far, no CWB has been unambiguously associated with TeV gamma-rays and only one has been detected in GeV gamma-rays \citep{2011A&A...526A..57F,2015A&A...577A.100R}. Wind blown bubbles from high mass stellar clusters and/or SNRs have long been proposed as sites of particle acceleration, through diffusive shock acceleration, up to PeV ($10^{15}$\,eV) energies \citep[e.g.][]{1980ApJ...237..236C} and recently, the first superbubble has been detected in TeV gamma-rays \citep{2015Sci...347..406H}. \citep{2014A&A...565A.118P} show that regions of high stellar activity within the central Galactic region, traced by dust emission and CO, could be correlated with TeV emission.

In summary, dense molecular gas that is both coincident with and/or adjacent to TeV emission can be very important in identifying the TeV emission mechanism and the location of particle accelerators. The transitions included in this study can not only help understand the density profile of molecular clouds, but can also identify regions of unusual astrophysical conditions such as outflows (e.g. the H$_2$O maser), shocked gas (e.g. the H$_2$O maser and NH$_3$ transitions) and ionised gas (e.g. H69$\alpha$). The kinematic velocity and linewidth of molecular transitions can provide information about the distance of and gas dynamics towards Galactic TeV gamma-ray sources. 

Previously, the $^{12}$CO(1-0) transition has been used to identify potential molecular cloud counterparts to unidentified TeV emission \citep[e.g.][]{2008A&A...483..509A, 2011A&A...531A..81H}. For example, by making a thorough comparison of two Galactic plane datasets, the HESS TeV gamma rays and the Nanten CO(1-0) emission, it was shown that the CO clouds in the W28 region have a remarkable spatial coincidence with the TeV gamma rays \citep{2008A&A...481..401A}. A subsequent comparative study toward the SNR RX\,J1713.7$-$3946 revealed that the combination of the atomic and molecular gas, i.e., the total interstellar hydrogen, shows a good spatial correspondence, where the Nanten CO J=1-0 distribution and the ATCA HI 21cm distribution were used to derive the gas column density \citep{2012ApJ...746...82F}. This correspondence is interpreted as indication for a hadronic component to the gamma rays from RXJ1713.7$-$3946 arising from the dense ISM clumps \citep{2010ApJ...708..965Z,2012ApJ...744...71I}. The interpretation is supported by the theoretical works on the VHE gamma rays which incorporates the highly inhomogeneous distribution of the interstellar medium \citep{2012ApJ...759...35I,2014MNRAS.445L..70G}. Further observational identification of the molecular and atomic gas toward the VHE sources are presented by \citet{2014ApJ...788...94F} for HESS\,J1731$-$347, and \citet{2012PASJ...64....8H} for HESS\,J1745$-$303. These observational and theoretical results indicate the important role of the interstellar medium whose density is in a wide range from 10 cm$^{-3}$ to 10$^5$ cm$^{-3}$ in producing the TeV gamma-rays, and warrant further systematic efforts to identify the interstellar gas toward the galactic HESS sources.

While $^{12}$CO is abundant, the $^{12}$CO(1-0) transition has a critical density of $\sim$1000\,cm$^{-3}$ which contributes to the transition becoming optically thick towards molecular cloud cores. In addition, CO rapidly depletes from the gas phase towards the centre of cloud cores \citep[e.g. see][]{2002ApJ...569..815T,2002ApJ...570L.101B}. Both of these aspects can hamper the understanding of molecular cloud density profiles and internal dynamics if the $^{12}$CO(1-0) transition is used on its own. Due to the coarse angular resolution of current TeV observations ($\sim0.1\degree$), the Dame CO survey has been used to trace large scale molecular content. As the angular and energy resolution of TeV gamma-ray observations improve with the new generation of ground based TeV telescope systems such as the Cherenkov Telescope Array (CTA, angular resolution 0.02$\degree$ to 0.2$\degree$), a better understanding of the density profiles of molecular clouds, using transitions which trace a wide range of temperatures and densities, will be needed to explore cosmic ray transport scenarios. Surveys such as MopraCO (angular resolution $\sim 0.01\degree$, \citet{2015PASA...32...20B}), Nanten2 (angular resolution $\sim 0.03\degree$) and H$_2$O Southern Galactic Plane Survey (HOPS) (angular resolution $\sim 0.03\degree$, \citet{2012MNRAS.426.1972P}), used here, will be invaluable for this cause.

Ideal tracers of dense gases such as NH$_3$, CS or HC$_3$N are widely used to trace molecular cloud cores due to their lower abundance (a factor of $\sim10^{-5} \times$ the CO abundance) and higher effective critical densities $\sim10^{4-5}$\,cm$^{-3}$. This gives them much lower optical depths in dense gas and so allows for a more accurate calculation of gas density and mass towards dense ($\sim10^{4-5}$\,cm$^{-3}$) molecular cloud cores. In addition, NH$_3$ depletes less rapidly than CO from the gas phase in cold molecular cloud cores \citep{2002ApJ...569..815T} and so is often used to trace cold ($< 30$K) gas which is often seen in infra red dark clouds (IRDCs). NH$_3$ inversion emission may be seen with collisional masers such as the H$_2$O maser traced in this study which is thought to trace protostellar jet termination shocks in early stages of star formation. The CS (1-0) and HC$_3$N transitions are used to trace warmer dense cores where star formation may already be switched on. These warmer cores are sometimes seen within H{\sc ii} regions, traced by radio recombination lines such as the H69$\alpha$ transition, where atomic hydrogen has been ionised by young, high mass stars.

The NH$_3$ molecule is a pyramidal symmetric top with inversion motion and metastable levels some of which display readily observable hyperfine structure. The NH$_3$\,(1,1) transition displays prominent satellite lines which along with NH$_3$\,(2,2) allow the optical depth, and hence gas temperature and mass to be strongly constrained \citep[e.g.][]{1983ARA&A..21..239H, 1983A&A...122..164W}. NH$_3$ has two distinct species, distinguished by the relative nuclear spins of the hydrogen atoms, ortho-NH$_3$ ($K = 3n$) in which all three spins are aligned and para-NH$_3$ (K $\neq 3n$) in which the three spins are not aligned. The rotational temperature within the same spin species reflects the kinetic temperature under Local Thermodynamic Equilibrium (LTE) conditions \citep{2009MNRAS.399..425M}. Since the transfer processes between ortho and para NH$_3$ are almost thermoneutral, conversion between the two spin species is very slow with respect to the reactive pathways (which are exothermic). The time scale of conversion processes between spin species is considered to be of the order of $10^6$ yr in the gas phase \citep{1969ApJ...157L..13C}. As a result, the "spin temperature" (or the rotational temperature between ortho and para species) is not considered to reflect the kinetic temperature of NH$_3$, but instead reflect formation conditions of the NH$_3$.

The ortho-NH$_3$ to para-NH$_3$ abundance ratio (OPR) is widely expected to be the statistical equilibrium value of 1.0 when the NH$_3$ molecules are formed in gas-phase or surface grain reactions \citep[e.g.][]{1999ApJ...525L.105U}. However, modelling by \citet{2013ApJ...770L...2F} suggests that, in gas phase reactions, the OPR can reasonably be expected to be $0.5 - 1.0$, and an OPR $<1$ reflects production from para-enriched H$_2$ gas. \citet{2013ApJ...770L...2F} have modelled the formation of ortho and para ammonia and have suggested that ortho ammonia is formed preferentially over para ammonia when the NH$_3$ is formed or condensed on a cold surface ($< 30$ K) such as water or ice. In addition, because the lowest energy level of the para species of NH$_3$ is 23 K higher than the lowest energy level of the ortho species, the para species require more energy for desorption than ortho species \citep{1999ApJ...525L.105U}. Both of these aspects would enhance the ortho-to-para abundance ratio (hereafter OPR) for NH$_3$ released from grain surfaces into the gas phase by shocks over the OPR of NH$_3$ produced in the gas phase. The observational study of \citet{1999ApJ...525L.105U} (amongst others) suggests that an enhanced OPR (OPR > 1) along with an enhanced NH$_3$ abundance can indicate a previous shock that has released NH$_3$ formed on dust grains into the gas phase.

This work provides a first systematic look at the dense gas (n >$10^4$\,cm$^{-3}$) traced by HOPS \citep{2012MNRAS.426.1972P} and further dedicated observations with the Mopra radio telescope towards Galactic TeV gamma-ray emission seen by H.E.S.S.. The large (8 GHz) bandwidth  of the Mopra spectrometer (MOPS) allows for the simultaneous observation of many molecular transitions. Multiple inversion-rotation transitions of NH$_3$ are included which allow us to identify molecular cloud cores displaying non-LTE conditions, which may indicate shocked gas. The velocity resolution ($\sim0.4$km/s at 22 GHz, \citealt{2010PASA...27..321U}) of MOPS at these wavelengths provides an opportunity to search for asymmetric line profiles of molecular transitions, which can indicate disruption and shocks within cloud cores. These Mopra observations are able to provide information about the densest parts of the interacting interstellar gas and will allow us to gain an insight in the cosmic ray penetration in the inner part of the cloud cores and the gamma-ray production.

\section{Observations and Data Reduction}
\label{observations}

Our study uses published H.E.S.S. results up to March 2015 (see Table \ref{table:detections} for specific references). H.E.S.S. detects TeV gamma-rays above an energy threshold of $\sim$100 GeV and up to $\sim$100 TeV with a typical energy resolution of 15\% per photon and an angular resolution of $\sim0.1\degree$ per event \citep{2006A&A...457..899A}. The H.E.S.S. field of view is $5\degree$ in diameter with a point source sensitivity of $\sim2.0\times10^{-13}$\,erg\,cm$^{-2}$\,s$^{-1}$ at 1 TeV (25 hrs obs).

\begin{figure*}
\includegraphics[width=0.95\textwidth]{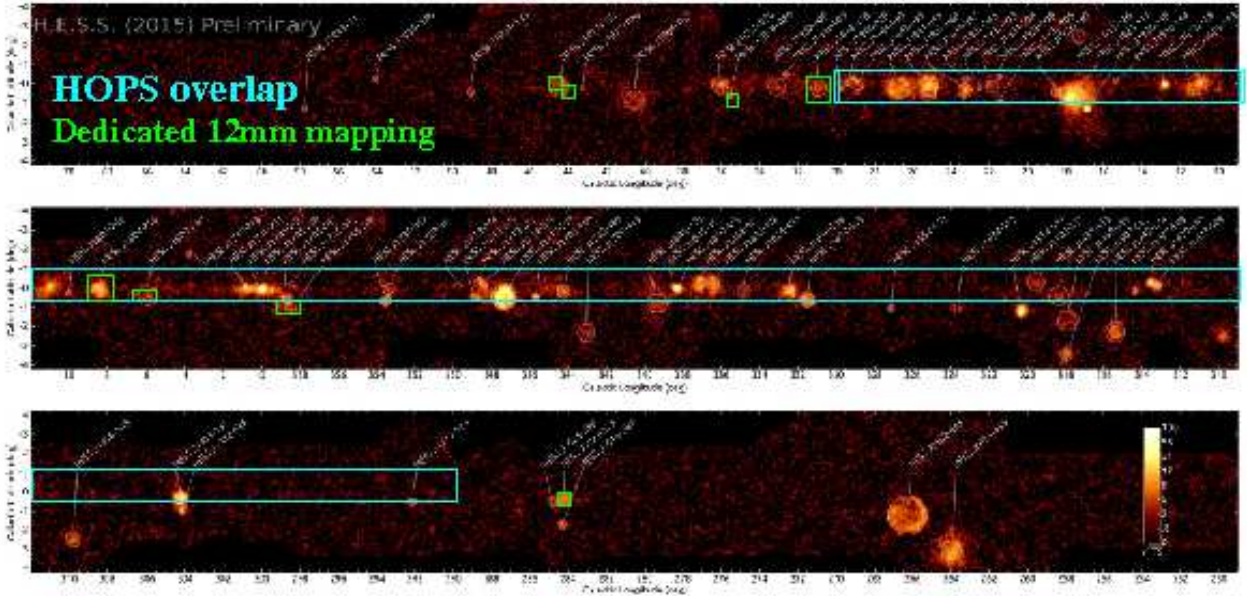}
\caption{\label{HESS_HOPS}TeV gamma-ray significance image taken from \citet{Proceedings:2015uxa} and overlaid with regions of HOPS 12\,mm coverage displayed using cyan boxes and our additional, dedicated 12\,mm observations displayed using green boxes.}
\end{figure*}

Molecular line data were taken from HOPS \citep{2011MNRAS.416.1764W} data and reduced using the ATNF packages LIVEDATA, GRIDZILLA, ASAP and MIRIAD (see http://www.atnf.csiro.au/computing/software/ for information on these packages). HOPS has mapped a 100 square degree strip of the Galactic plane (30$\degree$ > \textit{l} > -70$\degree$, $|b|$ < 0.5$\degree$) at 12\,mm wavelengths using the Mopra radio telescope. The telescope main beam size (FWHM) is $\sim 2$ arcmin at a wavelength of 12\,mm \citep{2010PASA...27..321U}. Mopra is a 22 m single-dish radio telescope located 450 km north-west of Sydney, near Coonabarabran, NSW, Australia. HOPS observations made use of the Mopra spectrometer (MOPS) in on-the-fly mapping mode. The zoom-mode of MOPS allows simultaneous observations from up to 16 spectral windows, where each window is 137.5 MHz wide and contains 4096 channels. HOPS targets water masers, thermal molecular emission and radio-recombination lines. Overlap of HOPS with the Galactic sky seen by H.E.S.S. can be seen in Fig. \ref{HESS_HOPS}. The Central Molecular Zone (CMZ) is not discussed in this study, and a study of this region will be detailed in a future paper. Further observation details of HOPS can be found described in \citet{2011MNRAS.416.1764W}. To derive kinematic distances from the local standard of rest (LSR) kinematic velocity, the Galactic rotation curve derived by \citet{1993A&A...275...67B} was used unless otherwise indicated.

We also carried out dedicated observations with the Mopra radio telescope between February 2011 and January 2015. These observations employed MOPS in zoom mode and utilised the same zoom bands and central frequency as HOPS. Five sites of deep ON/OFF pointing towards HESS\,J1801$-$233 and HESS\,J1745$-$303 were undertaken using modified zoom bands to include the NH$_3$\,(4,4) and (5,5) transitions. Deep ON/OFF pointing is where one beam-sized 'ON' region is observed, and calibrated using a single, beam-sized 'OFF region and are used in this study to provide extra sensitivity over mapping data. The FWHM of the Mopra beam within the 12\,mm band varies between 1.7 arcmin at the highest observed frequency in this data (27.5 GHz) and 2.4 arcmin at the lowest observed frequency (19.5 GHz) \citep{2010PASA...27..321U}. The mapping data, as with HOPS, uses MOPS in on-the-fly mapping mode with two scanning directions, Galactic longitude and latitude, in order to reduce scanning artefacts and noise. Each 0.5$\degree\times$0.5$\degree$ map took approximately 90 mins for each pass. Deep ON/OFF pointings were  were undertaken for 60 mins in selected regions where extra sensitivity was required. The mapping regions of our dedicated observations with HOPS-equivalent exposure have a mean T$_{\rm{rms}}$ of $0.2$ K per channel. Mapping regions with deeper coverage had four times the exposure, and so achieved a T$_{\rm{rms}}$\,$\sim$0.1 K per channel. Position-switched deep ON/OFF observations achieved a T$_{\rm{rms}}$ of $\geqslant\sim$0.02 K per channel. 
 
We applied the methods of \citet{1986A&A...157..207U} to estimate the NH$_3$ column density ($N_{\rm{NH}_3}$), the kinetic temperature (T$_{\rm{K}}$) and the H$_2$ density ($n_{\rm{H}_2}$) of the molecular clumps from the detected emission lines NH$_3$\,(J,K), J=K=1,2. The masses of molecular clumps are presented and discussed for six unidentified sources in detail in Section \ref{six_sources}. For clumps where all data are from HOPS the NH$_3$ gas parameters are presented in \cite{2012MNRAS.426.1972P}. Since our study is particularly interested in those unidentified TeV sources towards star forming regions, the TeV sources which are toward molecular clumps displaying H$_2$O maser emission and/or H$69\alpha$ emission are listed in Table. \ref{table:all_sources}. 

\subsection{TeV/Molecular Clump overlap}
Dense molecular clouds were included based on their emission in the NH$_3$\,(1,1) transition. The NH$_3$ data were smoothed over five velocity channels ($\sim2.0$\,km/s), improving the integrated S/N by a factor of $\sim\sqrt{5}$. We have classified molecular clouds as significant if their peak emission is >\,3$\sigma$, where $\sigma \equiv {\rm{T}_{\rm{rms}}}/{\sqrt{5}}$. Molecular clumps were identified as being towards the TeV emission if the clump centroid position fell within a 3$\sigma$ radius from the TeV centroid, and adjacent to the TeV emission if the clump centroid position fell between a 3 and 5$\sigma$ radius range from the TeV centroid, where $\sigma$ is the intrinsic rms of the TeV sources quoted in publication (relevant publications are noted in Table. \ref{table:detections}).

\begin{table*}
\caption{TeV sources covered by Mopra 12\,mm observations in our study. The 12\,mm observations include data from HOPS \citep{2011MNRAS.416.1764W} as well as additional observations undertaken specifically for our study. The H.E.S.S. TeV source name and centroid position (taken from given references) as well as some alternative names and whether NH$_3$\,(1,1) emission is detected towards, or adjacent to, the source is indicated. Sources highlighted in bold font are discussed in detail in this paper.}
\centering
\begin{threeparttable}[b]
\begin{tabular}{|c c c c c c c|}
\hline
H.E.S.S. name	    & Alternative name 	&\textit{l}	&\textit{b}	&	NH$_3$\,(1,1)	& NH$_3$\,(1,1)	&	TeV position/size\\ 
					&   				&			&			&	towards			& adjacent		&	reference	\\\hline
HESS\,J1023$-$575	& Westerlund 2 		&	284.22	&	$-$0.40	& 	- 				&	-			&	\tnote{1}	\\
HESS\,J1119$-$614	&					&	292.10	&	$-$0.49	&	-				&	-			&	\tnote{24}	\\
HESS\,J1303$-$631	&	             	&	304.24	&	$-$0.36	&	\checkmark	    &	-			&	\tnote{2}	\\ 
HESS\,J1418$-$609	& Kookaburra(Rabbit)&	313.25	&	0.15	&	-			   	&	\checkmark	&	\tnote{3}	\\ 
HESS\,J1420$-$607	& Kookaburra(PWN)	&	313.56	&	0.27	&	\checkmark	 	&	\checkmark	&	\tnote{3}	\\ 
HESS\,J1427$-$608	&					&	314.41	&	$-$0.14	&	\checkmark	 	&	\checkmark	&	\tnote{4}	\\ 
HESS\,J1457$-$593	& G318.2+0.1 		&	318.36	&	$-$0.43	&	-			 	&	\checkmark	&	\tnote{5}	\\ 
HESS\,J1503$-$582	&					&	319.62	&	0.29	&	-				&	-			&	\tnote{6}	\\ 
HESS\,J1614$-$518	&					&	331.52	&	$-$0.58	&	-			 	&	\checkmark	&	\tnote{4}	\\ 
HESS\,J1616$-$508	&					&	332.39	&	$-$0.14	&	\checkmark	 	&	\checkmark	&	\tnote{7}	\\ 
\textbf{HESS\,J1626$-$490}&				&	334.77	&	0.05	&	\checkmark	 	&	\checkmark	&	\tnote{4}	\\ 
HESS\,J1632$-$478	&					&	336.38	&	0.19	&	\checkmark	 	&	\checkmark	&	\tnote{4}	\\ 
HESS\,J1634$-$472	&					&	337.11	&	0.22	&	\checkmark	 	&	\checkmark	&	\tnote{4}	\\ 
\textbf{HESS\,J1640$-$465}&				&	338.32	&	$-$0.02	&	-			 	&	\checkmark	&	\tnote{7}	\\
HESS\,J1641$-$463	&					&	338.52	&	0.09	&	\checkmark 		&	-			&	\tnote{22}	\\		 
HESS\,J1646$-$458	& Westerlund 1		&	339.55	&	$-$0.35	&	\checkmark	  	&	\checkmark	&	\tnote{8}	\\ 
HESS\,J1702$-$420	&					&	344.30	&	$-$0.18	&	\checkmark	  	&	\checkmark	&	\tnote{4}	\\ 
HESS\,J1708$-$410	&   				&	345.68	&	$-$0.47	&	-			  	&	\checkmark	&	\tnote{4}	\\ 
HESS\,J1713$-$397	& RX J1713.7$-$3946	&	347.34	&	$-$0.47	&	\checkmark	  	&	\checkmark	&	\tnote{9}	\\ 
HESS\,J1713$-$381	& CTB 37B			&	348.65	&	0.38	&	-			 	&	-			&	\tnote{7}	\\ 
HESS\,J1714$-$385	& CTB 37A			&	348.39	&	0.11	&	-			  	&	-			&	\tnote{10}	\\ 
HESS\,J1718$-$385	&					&	348.83	&	$-$0.49	&	-				&	-			&	\tnote{11}	\\ 
\textbf{HESS\,J1729$-$345}&				&	353.44	&	$-$0.13	&	\checkmark	 	&	-			&	\tnote{12}	\\ 
HESS\,J1731$-$347	&					&	353.54	&	$-$0.67	&	-			  	&	\checkmark	&	\tnote{12}	\\ 
HESS\,J1741$-$302	&					&	358.4	&	0.19	&	-			 	&	-			&	\tnote{13}	\\ 
\textbf{HESS\,J1745$-$303}&			   	&	358.71	&	$-$0.64	&	\checkmark	 	&	\checkmark	&	\tnote{7}	\\ 
HESS\,J1800$-$240A	&					&	6.14 	& 	$-$0.63	&	\checkmark	 	&	-			&	\tnote{14}	\\ 
HESS\,J1800$-$240B	&					&	5.9		&	$-$0.37	&	\checkmark	 	&	-			&	\tnote{14}	\\ 
HESS\,J1800$-$240C	&					&	5.71	&	$-$0.06	&	-				&	\checkmark	&	\tnote{14}	\\ 
HESS\,J1801$-$233 	&	W28 			&	6.66	&	$-$0.27	&	\checkmark	 	&	-			&	\tnote{14}	\\ 
\textbf{HESS\,J1804$-$216}&				&	8.4		&	$-$0.03	&	\checkmark	 	&	-			&	\tnote{7}	\\ 
HESS\,J1808$-$204	&					&	9.96	&	$-$0.25	&	-				&	-			&	\tnote{23}	\\		
HESS\,J1809$-$193	&					&	11.18	&	$-$0.09	&	\checkmark	  	&	\checkmark	&	\tnote{11}	\\ 
HESS\,J1813$-$178	&					&	12.81	&	$-$0.03	&	-			  	&	\checkmark	&	\tnote{7}	\\ 
HESS\,J1818$-$154	& SNR G15.4$+$0.1	&	15.41	&	0.17	&	\checkmark	  	&	-			&	\tnote{15}	\\ 
HESS\,J1825$-$137	&					&	17.71	&	$-$0.7	&	-			  	&	\checkmark	&	\tnote{7}	\\ 
HESS\,J1828$-$099	&					&	21.49	&	0.38	&	- 				&	-			&	\tnote{22}	\\
HESS\,J1831$-$098	&   				&	21.85	&	$-$0.11	&	-			 	&	-			&	\tnote{16}	\\ 
HESS\,J1832$-$093	&					&	22.48	&	$-$0.16	&	\checkmark	  	&	-			&	\tnote{17}	\\ 
HESS\,J1832$-$085	&					&	23.21	&	0.29	&	-				&	-			&	\tnote{24}	\\
HESS\,J1834$-$087	&					&	23.24	&	$-$0.31	&	\checkmark	 	&	-			&	\tnote{7}	\\ 
HESS\,J1837$-$069	&					&	25.18	&	$-$0.12	&	\checkmark	 	&	\checkmark	&	\tnote{7}	\\ 
HESS\,J1841$-$055	&					&	26.8	&	$-$0.2	&	\checkmark	  	&	\checkmark	&	\tnote{4}	\\ 
HESS\,J1843$-$033	&					&	29.3	&	0.51	&	\checkmark	  	&	\checkmark	&	\tnote{18}	\\ 
HESS\,J1844$-$030	&					&	29.41	&	0.09	&	-				&	-			&	\tnote{24}	\\
HESS\,J1846$-$029	&	 				&	29.7	&	$-$0.24	&	-			 	&	-			&	\tnote{19}	\\ 
\textbf{HESS\,J1848$-$018}&				&	31.0	&	$-$0.16	&	\checkmark	 	&	-			&	\tnote{20}	\\ 
HESS\,J1858$+$020	&					&	35.58	&	$-$0.58	&	-		 	 	&	-			&	\tnote{4}	\\ 
HESS\,J1912$+$101	&					&	44.39	&	$-$0.07	&	\checkmark	 	&	-			&	\tnote{21} \\\hline
\end{tabular}
\centering
\label{table:detections} 
\begin{tablenotes}[justified]
\small{\item[1]\citet{2011A&A...525A..46H}\item[2]\citet{2005A&A...439.1013A}\item[3]\citet{2006A&A...456..245A}\item[4]\citet{2008A&A...477..353A}\item[5]\citet{2010tsra.confE.196H}\item[6]\citet{2008AIPC.1085..281R}\item[7]\citet{2006ApJ...636..777A}\item[8]\citet{2012A&A...537A.114A}\item[9]\citet{2006A&A...449..223A}\item[10]\citet{2008A&A...490..685A}\item[11]\citet{2007A&A...472..489A}\item[12]\citet{2011A&A...531A..81H}\item[13]\citet{2008AIPC.1085..249T}\item[14]\citet{2008A&A...481..401A}\item[15]\citet{Hofverberg:2011aa}\item[16]\citet{Sheidaei:2011vg}\item[17]\citet{2015MNRAS.446.1163H}\item[18]\citet{2008ICRC....2..579H}\item[19]\citet{2008ICRC....2..823D}\item[20]\citet{2008AIPC.1085..372C}\item[21]\citet{2008A&A...484..435A}\item[22]\citet{2014ApJ...794L...1A}\item[23]\citet{2012AIPC.1505..273R}\item[24]\citet{Proceedings:2015uxa}}
\end{tablenotes}
\end{threeparttable}
\end{table*}

The NH$_3$\,(1,1) transition and five further specific molecular transitions were searched for in our study towards the molecular clouds detected with NH$_3$\,(1,1) emission. The transitions were chosen based on their ability to trace star forming regions at various evolutionary phases and broad line gas within dense cloud cores. Three inversion-rotation transitions of NH$_3$, NH$_3$\,(J,K): J=K=1,2,3 were chosen in order to estimate gas temperature and density of molecular clumps as well as to identify regions of broad line emission, and unusual gas dynamics. The H$_2$O maser transition at 12\,mm was chosen to identify regions of ongoing star formation which may not show up in infra-red observations, and the radio recombination line H$69\alpha$ was chosen to identify regions of ionised gas where high mass stars have been recently formed. The cyanopolyyne transition HC$_3$N(3-2) was chosen as it, and other cyanopolyyne transitions, are thought to trace early stages of core evolution while NH$_3$ tends to become more abundant at later stages \citep{1992ApJ...392..551S}. The detections of these six molecular transitions are summarized in Table \ref{table:all_sources}. 

\section{Results overview}

Out of the 49 TeV sources included in our study, NH$_3$\,(1,1) emission was detected towards or adjacent to 38 of them (see Table \ref{table:detections} for a full list). Between one and nine molecular clumps are seen towards or adjacent to each TeV source. Around half of the molecular clumps detected in NH$_3$\,(1,1) emission display NH$_3$\,(2,2) emission and around one third display NH$_3$\,(3,3) emission. HC$_3$N (3-2) emission is detected in around one third of molecular clumps, as is the H$_2$O maser transition, while H$69\alpha$ emission is seen towards less than one fifth of molecular cores. These results are summarised in Table \ref{table:all_sources}. 

\begin{table*}
\caption{All molecular detections from molecular clumps with NH$_3$\,(1,1) emission towards Galactic TeV sources. This table shows an extract from the full table, which is available in the online appendix. Galactic coordinates of NH${_3}$(1,1) emission, with kinematic distance solutions obtained using the rotation curve of \citet{1993A&A...275...67B}, and coincident molecular emission lines.}
\centering
\begin{tabular}{|c c c c c c c c c c c|}
\hline
TeV Source	&	NH$_3$	&	NH$_3$	&	NH$_3$\,(1,1)				& kinematic & distance&	NH$_3$	&	NH$_3$	&	HC$_3$N	&	H$_2$O 	&	H69$\alpha$\\
			&	(1,1)	&	(1,1)	&\textit{V$_{\rm{LSR}}$}	&near&far&	(2,2)	&	(3,3)	&	(3-2)	&	(6-5)	&		\\
			&\textit{l}	&\textit{b}	&	 (km/s)					&(kpc)&(kpc)&			&			&			&			&		\\\hline

HESS\,J1729$-$345	&	353.3	&	$-$0.1	&	$-$16.1	&3.8&10.3&	-	&	-	&	-	&	-	&	-	\\
HESS\,J1731$-$347	&	353.4	&	$-$0.3	&	$-$18.7	&4.3&9.8&	\checkmark	&	\checkmark	&	\checkmark	&	-	&	-	\\
HESS\,J1745$-$303	&	358.4	&	$-$0.5	&	6.5		& - & - &	\checkmark	&	-	&	-	&	\checkmark	&	-	\\
					&	358.5	&	$-$0.4	&	$-$3.7	&2.6&10.5&	\checkmark	&	\checkmark	&	-	&	\checkmark	&	-	\\
					&	358.6	&	$-$0.8	&	2.7		&19.5& - &	\checkmark	&	-	&	\checkmark	&	-	&	-	\\
					&	358.6	&	$-$0.4	&	$-$6.7	&6.3& - &	-	&	-	&	-	&	-	&	-	\\
					&	358.8	&	$-$0.4	&	$-$29.7	& - & - &	\checkmark	&	\checkmark	&	\checkmark	&	-	&	-	\\
					&	358.8	&	$-$0.5	&	$-$54.0	& - & - &	\checkmark	&	\checkmark	&	\checkmark	&	-	&	-	\\	
HESS\,J1800$-$240C	&	5.6		&	$-$0.1	&	$-$27.2	& - & - &	\checkmark	&	\checkmark	&	\checkmark	&	\checkmark	&	-	\\
					&	5.8		&	$-$0.2	&	12.0	&3.5&10.7&	-	&	-	&	-	&	-	&	-	\\
HESS\,J1800$-$240B	&	5.9		&	$-$0.4	&	7.7		&2.3&11.9&	\checkmark	&	\checkmark	&	\checkmark	&	\checkmark	&	\checkmark	\\
					&	5.9		&	$-$0.3	&	9.0		&2.6&11.5&	\checkmark	&	-	&	\checkmark	&	-	&	-	\\
HESS\,J1800$-$240A	&	6.6		&	$-$0.3	&	6.0		&1.6&12.6&	\checkmark	&	\checkmark	&	-	&	-	&	-	\\
					&	6.8		&	$-$0.3	&	20.1	&4.7&9.3&	\checkmark	&	-	&	\checkmark	&	\checkmark	&	-	\\
HESS\,J1804$-$216	&	8.1		&	0.2		&	18.0	&3.7&10.5&	\checkmark 	&	\checkmark	&	-	&	-	&	\checkmark	\\
					&	8.3		&	0.2		&	17.1	&3.5&10.7&	\checkmark	&	- 	&	-	&	-	&	-	\\
					&	8.4		&	$-$0.3	&	36.7	&6.7& - &	\checkmark	&	\checkmark	&	\checkmark	&	-	&	-	\\
					&	8.7		&	$-$0.4	&	36.7	&6.4& - &	\checkmark	&	\checkmark	&	\checkmark	&	\checkmark	&	-	\\\hline

\end{tabular}
\label{table:all_sources} 
\end{table*}

\subsection{NH$_3$ Linewidths}

The FWHM of the NH$_3$ main line is often used as a measure of the total energy (thermal plus non-thermal) associated with a molecular clump. Broader lines are expected from regions with higher temperatures or additional dynamics (due to turbulence, infall, outflow or tidal flow). The purely Maxwell-Boltzmann thermal linewidth FWHM $\Delta\,v_{th}$ expected from a gas at temperature \textit{T} is given by:
\begin{equation}
\centering
\Delta\,v_{th} = \sqrt{\frac{8\ln(2)kT}{m_{\rm{NH}_3}}} \rm{km s}^{-1} 
\label{eq:linewidth}
\end{equation}
where \textit{k} is Boltzmann's constant and $m_{\rm{NH}_3}$ is the mass of the NH$_3$ molecule. For a temperature of 15 K, as might be expected in typically cold, dense NH$_3$ cores, a thermal linewidth of 0.20 km.s$^{-1}$ is obtained, \citep{1983ARA&A..21..239H}. The line FWHM, of each core was estimated from a Gaussian fit to the main peak of the emission with additional Gaussians to fit each of the four satellite lines, which are generally resolved in the (1,1) transition. 

The FWHM for all molecular clumps identified in our study is noticeably wider than that expected from purely thermal broadening, suggesting additional non-thermal or kinetic energy which dominates over broadening from the instrumental response, which in zoom mode at 12\,mm provides a velocity resolution of $\sim0.4$\,km/s \citep{2010PASA...27..321U}. Typical linewidth FWHMs observed for NH$_3$ emission in our study are from 1-2 km/s, but in a few cases, very broad linewidths (> 10km/s) are observed. For NH$_3$\,(1,1) emission with very broad linewidths, the satellite lines are blended with the main line which makes fitting the five Gaussian components non-trivial. For these sources, the NH$_3$\,(2,2) linewidth (assumed to be unaffected by satellite line blending) was used as an upper limit to the width of the NH$_3$\,(1,1) main line, and a temperature and density calculated based on this fit. The TeV sources towards which these very broad linewidths are found, HESS\,J1801$-$233 (the TeV source to the NE of the SNR W28 towards a SNR/molecular cloud interaction region) \citep{2011MNRAS.411.1367N} and HESS\,J1745$-$303 \citep{2008A&A...483..509A}, are both thought to be produced hadronically through interaction of accelerated protons and dense gas. The broad linewidths observed towards these sources, together with coincident broad SiO(1-0) emission (which can be seen in \cite{2012MNRAS.419..251N} for HESS\,J1801$-$233 and in an upcoming paper for HESS\,J1745$-$303), indicates the previous passage of a shock through this gas.

\subsection{Ortho-to-para NH$_3$ Abundance Ratios} \label{ortho-para}

Para NH$_3$\,(1,1) and NH$_3$\,(2,2) detections (as well as ortho NH$_3$\,(3,3) detections when available) and the LTE methods summarised in \citet{1986A&A...157..207U} were used to determine the rotational temperatures of all molecular clumps towards TeV sources in the HOPS dataset. Towards HESS\,J1745$-$303 and HESS\,J1801$-$233 where NH$_3$\,(3,3) detections were made, it was noticed when calculating rotational temperatures that the optical depths did not decrease with increasing J (as would be expected) and the NH$_3$\,(3,3) brightness temperature > NH$_3$\,(1,1) brightness temperature which is shown in Figs \ref{J1745_33_to_11_image} and \ref{W28_33_to_11_ratio_image} as well as the spectra in Figs \ref{J1745_NH3_spectra_REG1}, \ref{J1745_NH3_spectra_REG2}, \ref{J1745_NH3_spectra_REG4}, \ref{W28_NH3_spectra_REG1} and \ref{W28_NH3_spectra_REG2}. Subsequently, in this work, where NH$_3$\,(3,3) brightness temperature > NH$_3$\,(1,1) brightness temperature, the NH$_3$\,(J,K), J=K > 2 emission is not expected to represent the same excitation temperature as the NH$_3$\,(1,1) transition and so it is assumed that the NH$_3$\,(J,K) J=K=3,4,5,6 emission is optically thin. The molecular clumps towards HESS\,J1745$-$303 and HESS\,J1801$-$233 with NH$_3$\,(3,3) brightness temperature > NH$_3$\,(1,1) brightness temperature were found to have extended emission in all four of the metastable ammonia transitions considered in the HOPS observations (NH$_3$\,(J,K) J=K=1,2,3,6). The molecular clouds towards both {HESS\,J1745$-$303 and HESS\,J1801$-$233 exist close to supernova remnants (SNRs) that are interacting with molecular clouds, evident by 1720 MHz OH masers. In addition, 7 mm observations of the regions displaying NH$_3$\,(3,3)-to-NH$_3$\,(1,1) brightness temperature ratios > 1 revealed extended SiO\,(1-0) emission which matches the morphology of the NH$_3$ emission. 

Upper state column densities for each transition were calculated using the LTE analysis summarised by \citet{1986A&A...157..207U}. These values were divided by the total degeneracy (angular momentum degeneracy, g$_u = 2J + 1$ as well as spin and K degeneracy) and used to plot rotation diagrams. Rotation diagrams shown in Figs \ref{1745_region1_Nuplot}, \ref{1745_region2_Nuplot}, \ref{1745_region3_Nuplot}, \ref{W28_region1_Nuplot} and \ref{W28_region2_Nuplot} include upper-state column densities of each observed transition towards each region of extended NH$_3$ emission where NH$_3$\,(3,3) brightness temperature $\geqslant$ NH$_3$\,(1,1) brightness temperature. The rotation temperature of para NH$_3$ was estimated separately for the NH$_3$\,(J,K) J=K=1,2 transitions and the NH$_3$\,(J,K) J=K=4,5 transitions. The rotational temperature of ortho NH$_3$ was estimated using the NH$_3$\,(J,K) J=K=3,6 transitions, and it is assumed that T$_{\rm{36}} =$ T$_{\rm{45}}$.

ON/OFF deep pointing observations were taken at three positions towards HESS\,J1745-303, and two positions towards HESS\,J1801$-$233. These regions indicated in Fig. \ref{J1745_33_to_11_image}, which were chosen due to their high NH$_3$ (3,3)-to-(1,1) brightness temperature ratios, and Fig. \ref{W28_33_to_11_ratio_image} which were chosen due to their positions immediately post shock and pre shock where the shock position is traced by 1720 MHz OH masers. The method of \citet{1999ApJ...525L.105U} was used to estimate the ortho-to-para NH$_3$ abundance ratio. Figs \ref{1745_region1_Nuplot}, \ref{1745_region2_Nuplot}, \ref{1745_region3_Nuplot}, \ref{W28_region1_Nuplot} and \ref{W28_region2_Nuplot} display a straight line fit to the derived NH$_3$\,(4,4) and (5,5) column densities, where 1/slope = T$_{\rm{rot}}$. This rotational temperature was compared to the T$_{\rm{rot}}$ fit for the ortho NH$_3$\,(3,3) and (6,6), and in many cases the lines fitting each pair of column densities is almost parallel (indicating an equivalent T$_{\rm{rot}}$). Ortho-to-para NH$_3$ abundance ratios were estimated and used to produce reduced ortho NH$_3$\,(3,3) and (6,6) column densities ($\textit{ROCD}$s) according to 
\begin{equation}
\textit{ROCD} = \textit{N}_{ortho}/\rm{OPR}_{est} 
\label{equation:reduced_ortho}
\end{equation}
where $\textit{N}_{ortho}$ is the upper state ortho NH$_3$ column density for each transition, and $\rm{OPR}_{est}$ is the estimated NH$_3$ ortho-to-para abundance ratio. These 'reduced' ortho NH$_3$ column densities were then optimised with least squares fitting with a straight line fit to all NH$_3$\,(J,K) J=K=3,4,5,6 column densities. The ortho-to-para abundance ratio which gives the best fit is given in Table. \ref{table:OPR}. 

\begin{table*}
\centering
\begin{tabular}{|c c c c c c|}
\hline
TeV Source			&T$_{rot}$	&T$_{rot}$	&T$_{rot}$	&	T$_{rot}$	& 					\\
and region			&	para	&	ortho	&	para 	&(3,3)-(6,6)	& OPR				\\
					&(1,1),(2,2)&(3,3),(6,6)&(4,4),(5,5)&(reduced ortho)& best fit			\\
					&(K)		&(K)		&(K)		&(K)			&					\\\hline
					
HESS\,J1745$-$303	&			&			&			& 				& 					\\
region 1			&	68		&	222		&	239		&	224 		&$1.5\pm0.3$		\\
region 2			&	64		&	256		&	232		&	234			&$1.8\pm0.6$		\\
region 3			&	91		&	247		&	181		&	239			&$1.7^{+0.3}_{-0.4}$\\\hline
HESS\,J1801$-$233	&			&			&			& 				& 					\\
region 1			&	108		&	285		&	360		&	291 		&$1.5\pm0.5$		\\
region 2			&	143		&	259		&	-		&	-			&> 2.0*				\\\hline

\end{tabular}
\caption{The T$_{rot}$ given for NH$_3$\,(3,3)-(6,6) (reduced ortho) = 1/(slope of the line fit to the column densities for regions defined in Figs. \ref{J1745_33_to_11_image} and \ref{W28_33_to_11_ratio_image} (adjusted according to the estimated NH$_3$ Ortho-to-Para abundance Ratio (OPR))) of the transitions for NH$_3$\,(J,K) J=K=3,4,5,6. The OPR given is that with the best fit to the NH$_3$\,(J,K) J=K=3,4,5,6 column densities. The * indicates a region where the apparent OPR fitted appears to be due to an incorrect division of transitions into temperature components rather than an enhanced OPR. Further details are outlined in the caption of Fig. \ref{W28_region2_Nuplot}.}
\label{table:OPR} 
\end{table*}

It can be see in Table. \ref{table:OPR} that for all three regions investigated towards HESS\,J1745-303, a an OPR $\geqslant$ 1 is indicated (see Table. \ref{table:OPR}). We believe an OPR $\geqslant$ 1 indicates the previous passage of a shock through a gas cloud (see Section \ref{introduction}). The regions observed towards HESS\,J1745$-$303 with an OPR $\geqslant$1 lie outside the boundary of the SNRs in the region indicating that another shock is responsible for the enhancement of ortho-NH$_3$ over para-NH$_3$. Several radio continuum point sources, centred on a molecular CO (1-0) ring in the region, may be remnants of an OB association responsible for a superbubble \citep{1992ApJ...398..128U}.

\begin{figure}
\centering
\includegraphics[width=0.5\textwidth, angle=0]{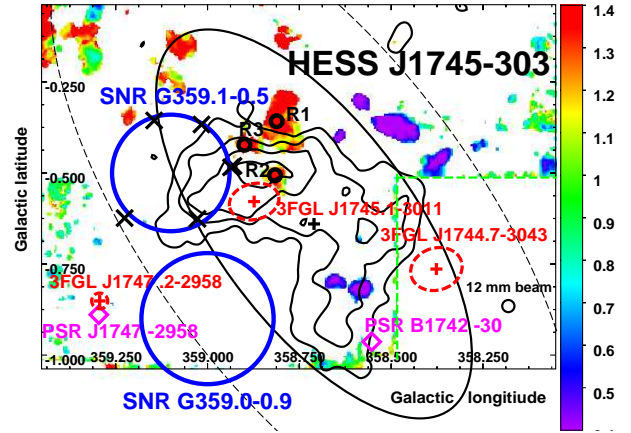}
\caption{An image of NH$_3$\,(3,3) peak pixel-to-NH$_3$\,(1,1) peak pixel emission towards TeV source HESS\,J1745-303. The image has been produced to only show the ratio in regions of significant NH$_3$\,(3,3) emission although some scanning artefacts from noisy pixels are still visible towards the galactic East and South edges of the mapping region. This image is overlaid with contours of TeV emission \textit{(black)} and the regions towards which the deep pointing spectra were taken \textit{(small black circles)}. Supernova remnants are indicated in large, solid blue circles, 1720 MHz OH masers are indicated with black $\times$s and pulsars are indicated in magenta diamonds.}
\label{J1745_33_to_11_image}
\end{figure}

\begin{figure}
\centering
\includegraphics[width=0.5\textwidth, angle=0]{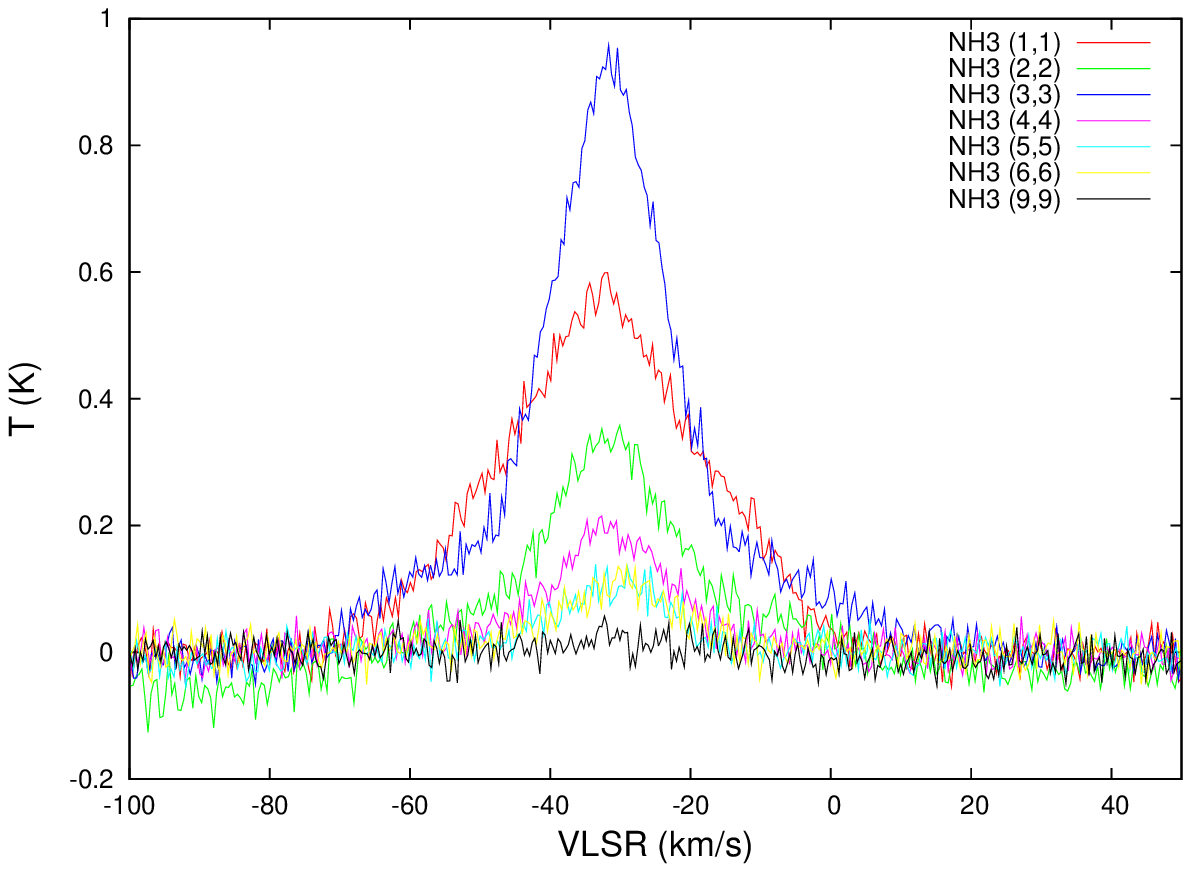}
\caption{NH$_3$\,(J,K) J=K=1,2...6 spectra from the region marked as region 1 towards HESS\,J1745-303 in Figure \ref{J1745_33_to_11_image}. The peak brightness temperature of the NH$_3$\,(3,3) emission is greater than that of the other transitions.}
\label{J1745_NH3_spectra_REG1}
\end{figure}

\begin{figure}
\centering
\includegraphics[width=0.5\textwidth, angle=0]{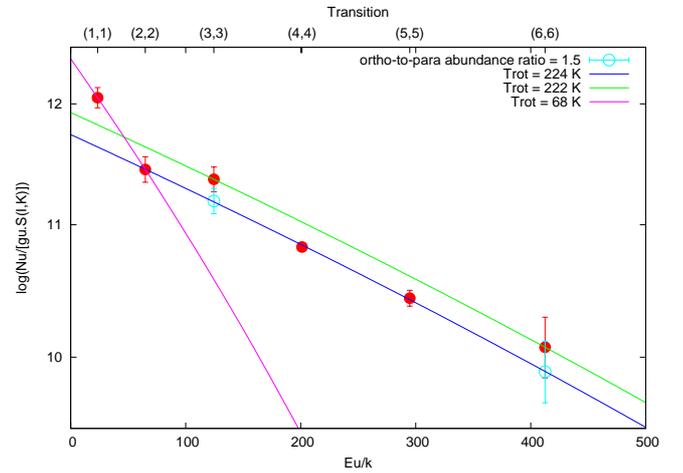}
\caption{Rotation diagram for region 1 of HESS\,J1745$-$303 (shown in Fig.  \ref{J1745_33_to_11_image}) where N$_u$, g$_u$, and E$_u$ are the column density, the statistical weight, and the energy for the upper levels of the transitions NH$_3$\,(J,K) J=K=1,2,...6. Rotational temperatures, T$_{\rm{rot}} =$ 1/slope, have been determined for three groups of transitions, and it can be seen, since the slopes are very similar, that the rotational temperature is similar for the ortho NH$_3$\,(3,3) and (6,6) transitions and the para NH$_3$\,(4,4) and (5,5) for which the rotational temperature given here was fit along with the OPR adjusted ortho column densities (shown as open cyan circles). Here an OPR of 2.0 was estimated. The best fit OPR is $1.5\pm{0.5}$.}
\label{1745_region1_Nuplot}
\end{figure}

\begin{figure}
\centering
\includegraphics[width=0.5\textwidth, angle=0]{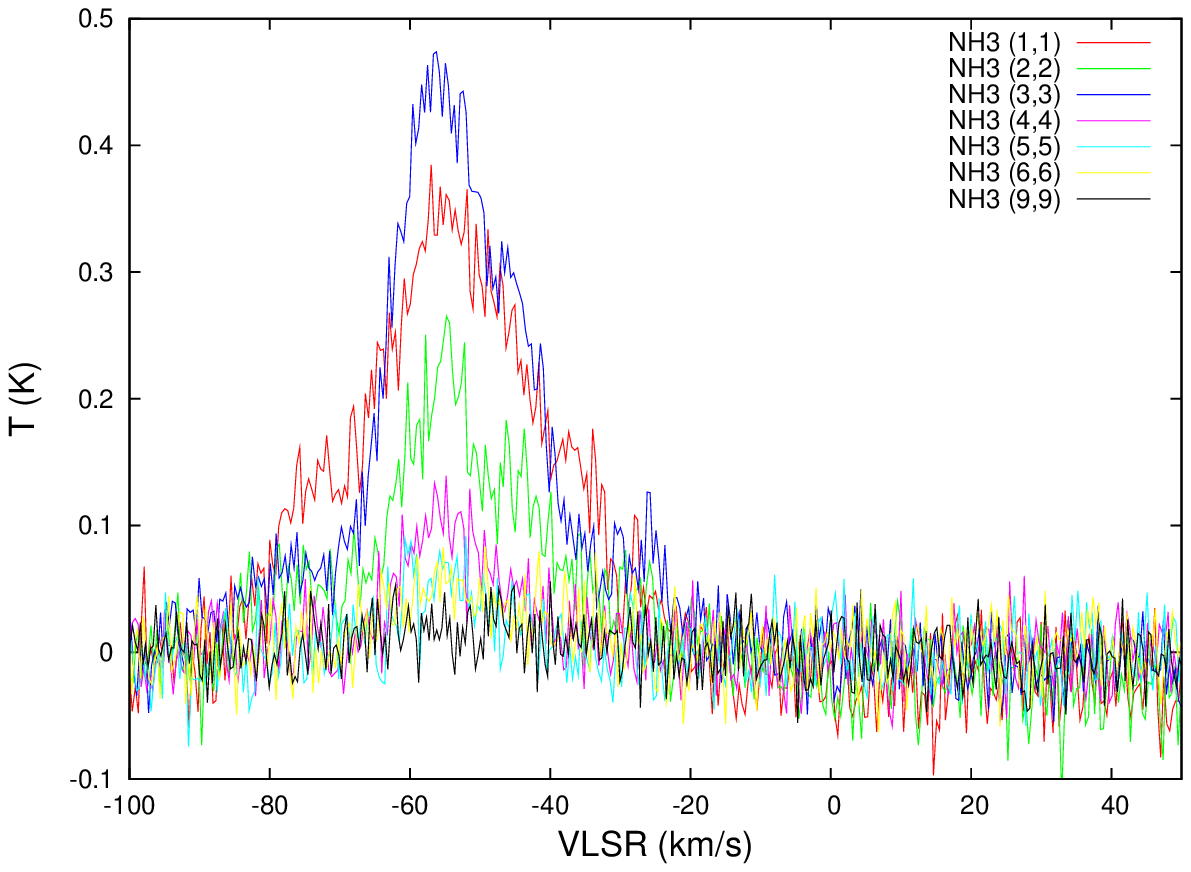}
\caption{NH$_3$\,(J,K) J=K=1,2...6 spectra from the region marked as region 2 towards HESS\,J1745-303 in Figure \ref{J1745_33_to_11_image}. The peak brightness temperature of the NH$_3$\,(3,3) emission is greater than that of the other transitions.}
\label{J1745_NH3_spectra_REG2}
\end{figure}

\begin{figure}
\centering
\includegraphics[width=0.5\textwidth, angle=0]{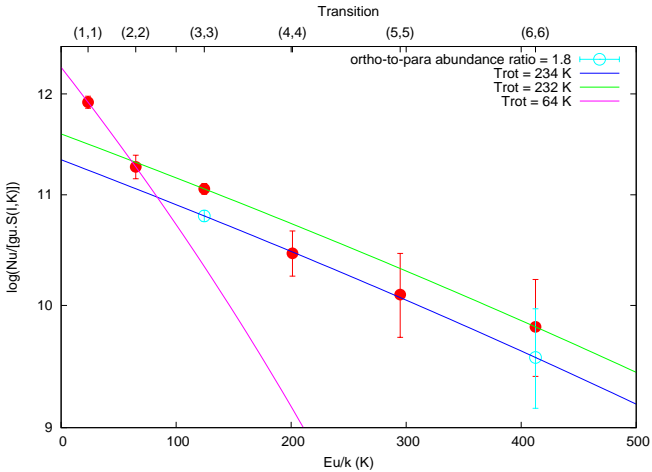}
\caption{As in Fig. \ref{1745_region1_Nuplot}, for region 2 of HESS\,J1745$-$303 (shown in Fig. \ref{J1745_33_to_11_image}). For this region an OPR of 2.0 was estimated. The best fit OPR is $1.8\pm{0.6}$, indicating an OPR enhancement. This region lies towards the highest significance of TeV gamma-ray emission.}
\label{1745_region2_Nuplot}
\end{figure}

\begin{figure}
\centering
\includegraphics[width=0.5\textwidth, angle=0]{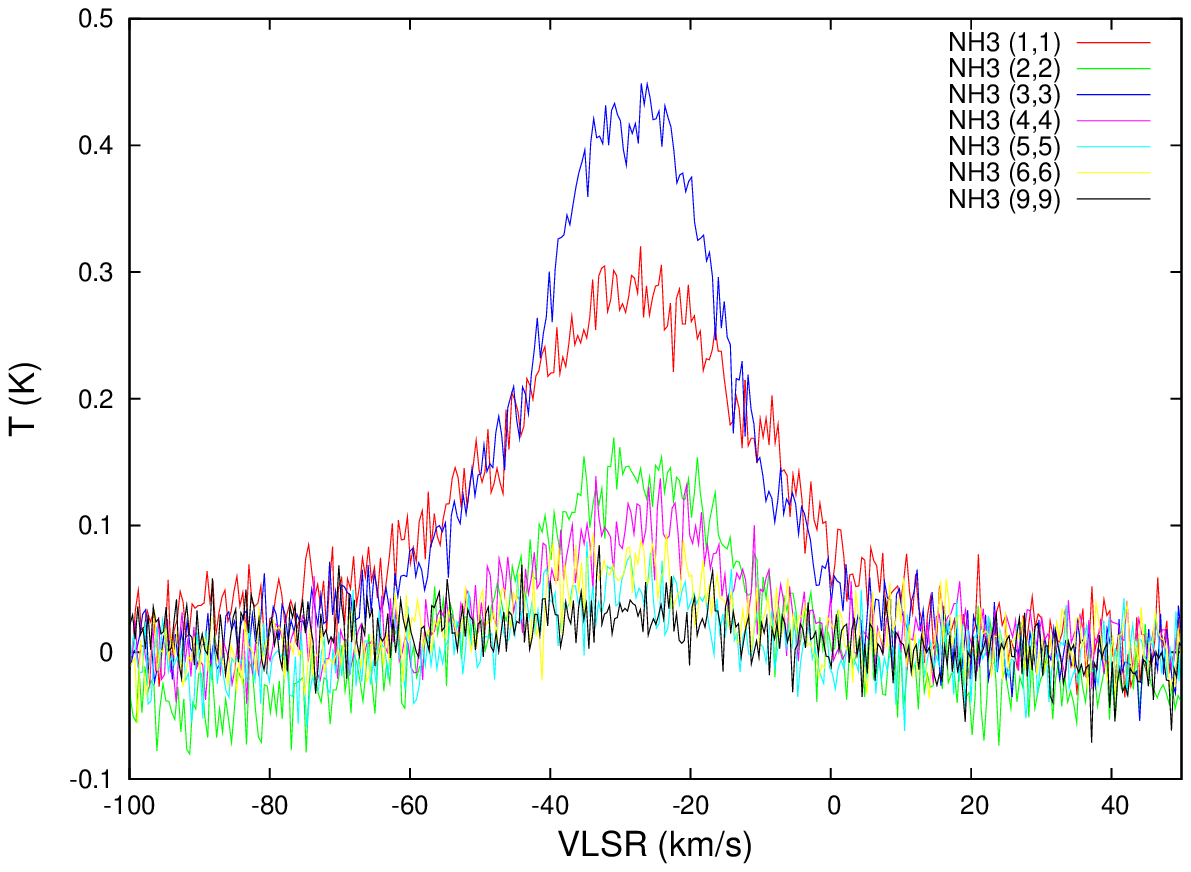}
\caption{NH$_3$\,(J,K) J=K=1,2...6 spectra from the region marked as region 3 towards HESS\,J1745-303 in Figure \ref{J1745_33_to_11_image}. The brightness temperature of the NH$_3$\,(3,3) emission is greater than that of the other transitions.}
\label{J1745_NH3_spectra_REG4}
\end{figure}

\begin{figure}
\centering
\includegraphics[width=0.5\textwidth, angle=0]{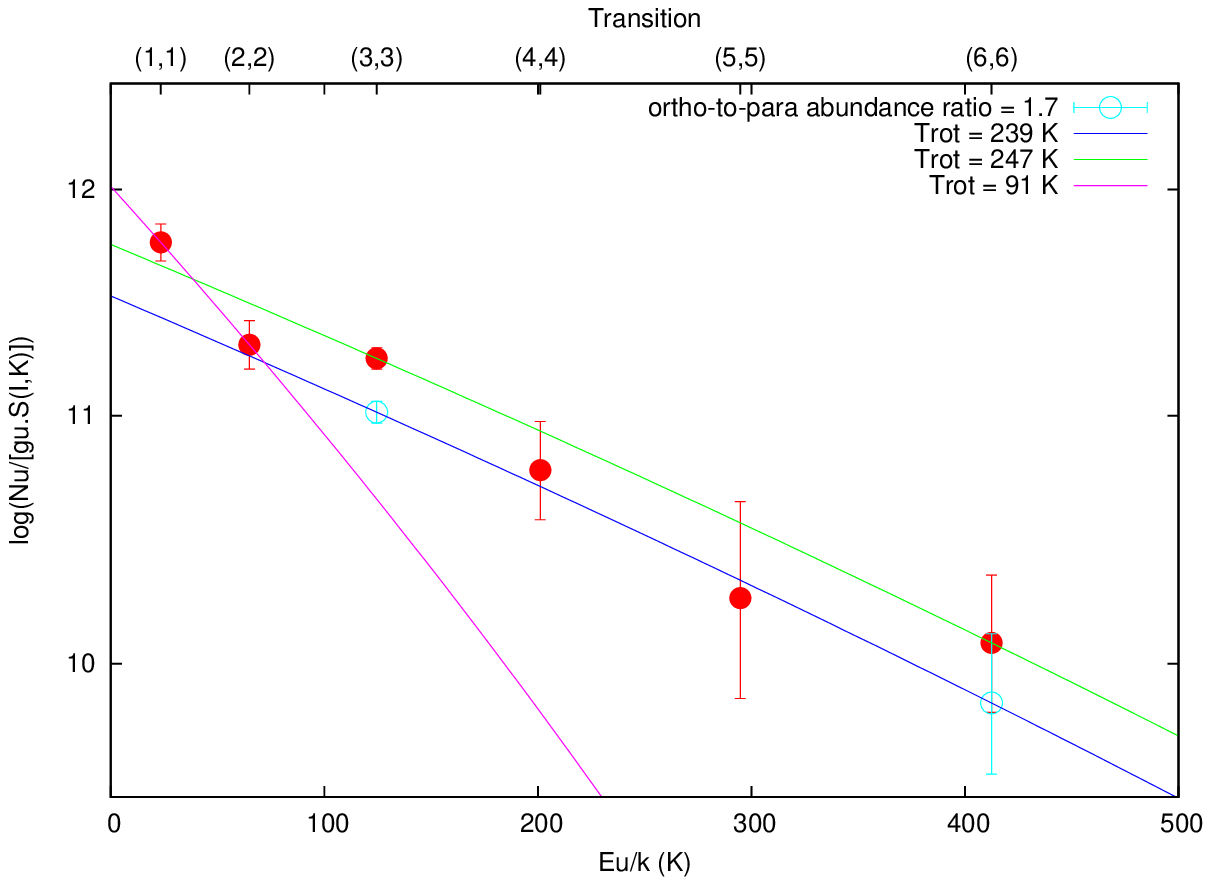}
\caption{As in Fig. \ref{1745_region1_Nuplot}, for region 3 of HESS\,J1745$-$303 (shown in Fig. \ref{J1745_33_to_11_image}). For this region an OPR of 1.5 was estimated. The best fit OPR is $1.7^{+0.3}_{-0.4}$.}
\label{1745_region3_Nuplot}
\end{figure}

ON-OFF deep pointing observations were taken at two positions towards HESS\,J1801$-$233. These regions are indicated in Fig. \ref{W28_33_to_11_ratio_image} and were chosen due to their positions immediately post shock and pre shock where the shock position is traced by 1720 MHz OH masers. 

For HESS\,J1801$-$233, the method used here cannot determine whether the OPR is enhanced in the regions indicated in Fig. \ref{W28_33_to_11_ratio_image}. Region 1 is immediately post shock and region 2 is immediately pre-shock where the shock position is traced by 1720 MHz OH masers. The best fit indicates that the OPR is enhanced in both regions which is consistent with a previous study \citep{2016MNRAS.462..532M} using NH$_3$ (J,K), J=K=1,2,3,4,6 to estimate the OPR in region 1. For region 2 (indicated in Fig. \ref{W28_33_to_11_ratio_image}) the rotational temperature for NH$_3$ (1,1) and (2,2) fits to the column densities of transitions up to the NH$_3$ (4,4) upper limit (as can be seen in Fig. \ref{W28_region2_Nuplot}), which may indicate that the higher J NH$_3$ (5,5) and (6,6) transitions are tracing a higher temperature component of the gas. If this were the case, the lower limit OPR for this region would not reflect the conditions of the gas. An enhanced OPR would not be expected in this region. In region 1 (indicated in Fig. \ref{W28_33_to_11_ratio_image}), the post shocked region, the determined OPR is consistent with 1, within errors, however an enhanced OPR could be expected in this region. 

\begin{figure}
\centering
\includegraphics[width=0.5\textwidth, angle=0]{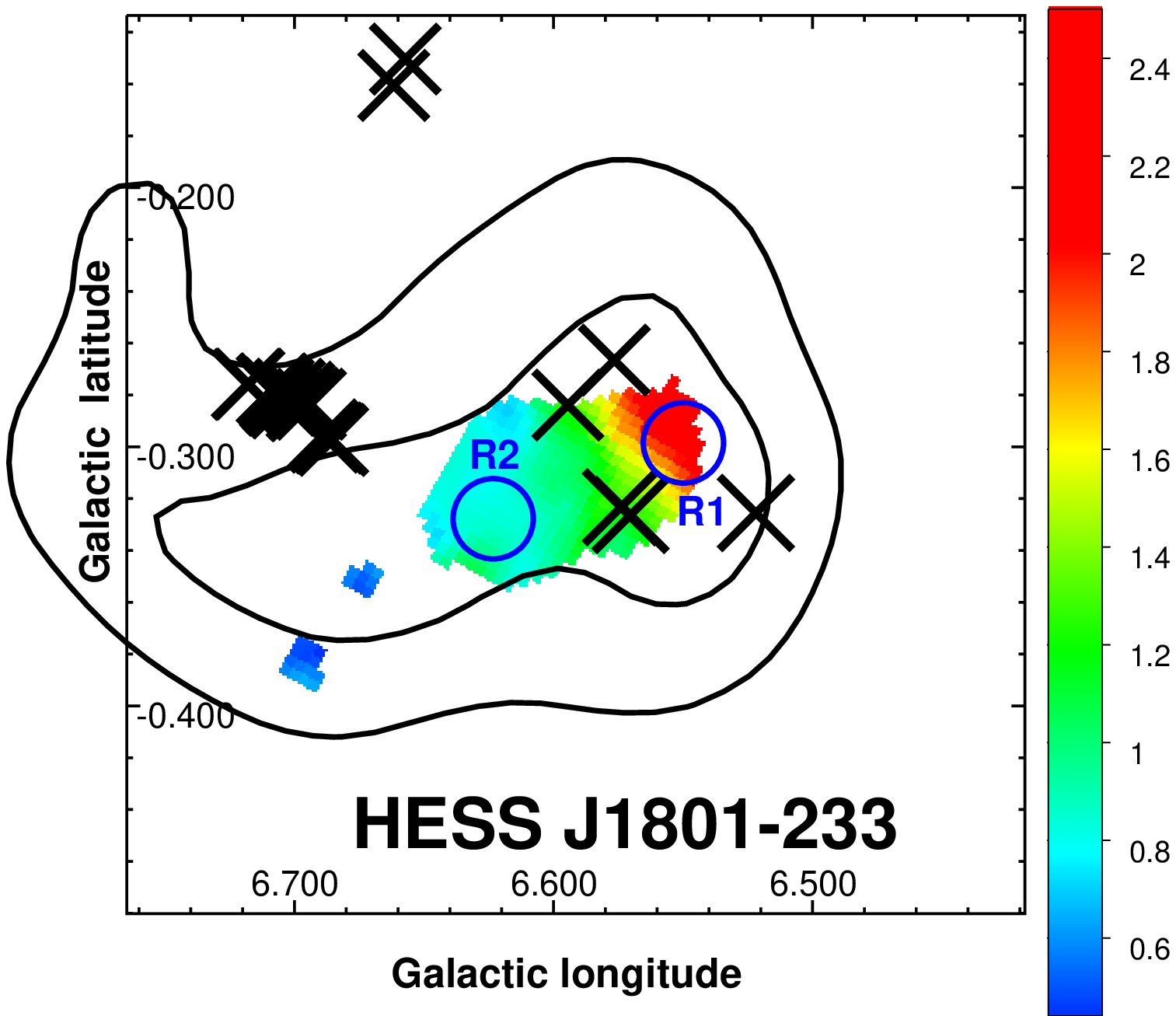}
\caption{An image of NH$_3$\,(3,3) peak pixel-to-NH$_3$\,(1,1) peak pixel emission towards TeV source HESS\,J1801$-$233. This image is overlaid with contours of TeV emission \textit{(black)} and the regions towards which the deep pointing spectra were taken \textit{(blue circles)}. 1720 MHz OH masers are indicated with black $\times$s which occur immediately post-shock of the SNR. The SNR shock here is moving from R1 towards R2.}
\label{W28_33_to_11_ratio_image}
\end{figure}

\begin{figure}
\centering
\includegraphics[width=0.5\textwidth, angle=0]{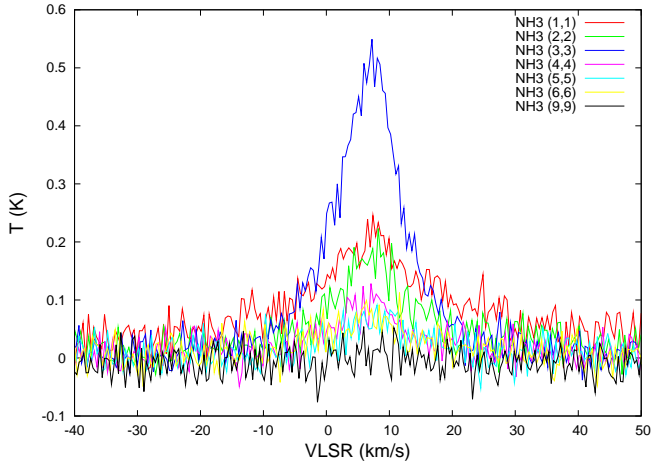}
\caption{NH$_3$\,(J,K) J=K=1,2...6 spectra from the region marked as R1, towards HESS\,J1801$-$233, in Figure \ref{W28_33_to_11_ratio_image}. This region is post shock as the 1720 MHz OH masers effectively mark the current position immediately post shock, and the SNR is expanding with the shock moving towards R2. These spectral lines all display broad (> 6 km/s) linewidths and this region also displays extended SiO (1-0) emission. The brightness temperature of the NH$_3$\,(3,3) emission is greater than that of the other transitions.}
\label{W28_NH3_spectra_REG1}
\end{figure}

\begin{figure}
\centering
\includegraphics[width=0.5\textwidth, angle=0]{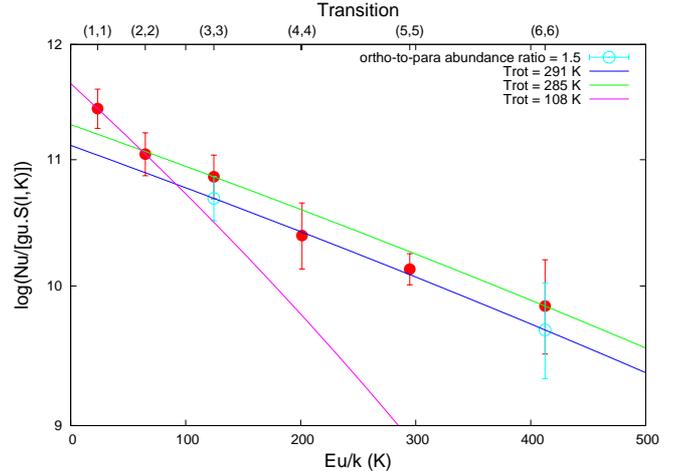}
\caption{As in Fig. \ref{1745_region1_Nuplot}, for R1 of HESS\,J1801$-$233 (shown in Fig. \ref{W28_33_to_11_ratio_image}). For this region an OPR of 1.5 was estimated. The best fit OPR is $1.5\pm{0.5}$. This region is the post shocked region.}
\label{W28_region1_Nuplot}
\end{figure}

\begin{figure}
\centering
\includegraphics[width=0.5\textwidth, angle=0]{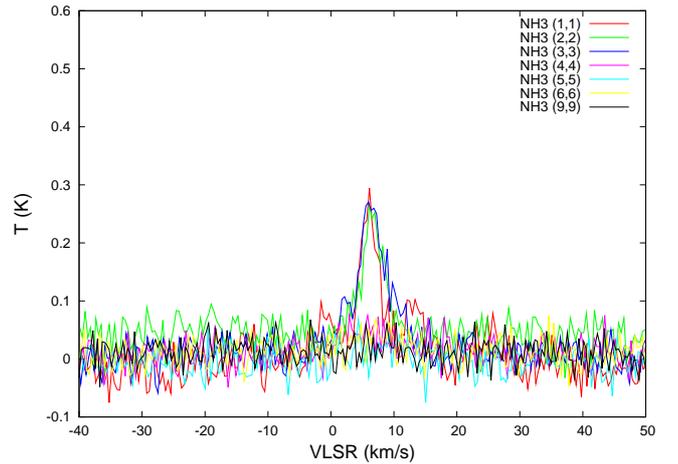}
\caption{NH$_3$\,(J,K) J=K=1,2...6 spectra from the region marked as R2, towards HESS\,J1801$-$233, in Figure \ref{W28_33_to_11_ratio_image}. This region is pre shock as the 1720 MHz OH masers effectively mark the current position immediately post shock, and the SNR is expanding with the shock moving towards R2. The brightness temperature of the NH$_3$\,(3,3) emission is slightly less than that of the NH$_3$\,(1,1) emission and slightly greater than that of the NH$_3$\,(2,2) emission.}
\label{W28_NH3_spectra_REG2}
\end{figure}

\begin{figure}
\centering
\includegraphics[width=0.5\textwidth, angle=0]{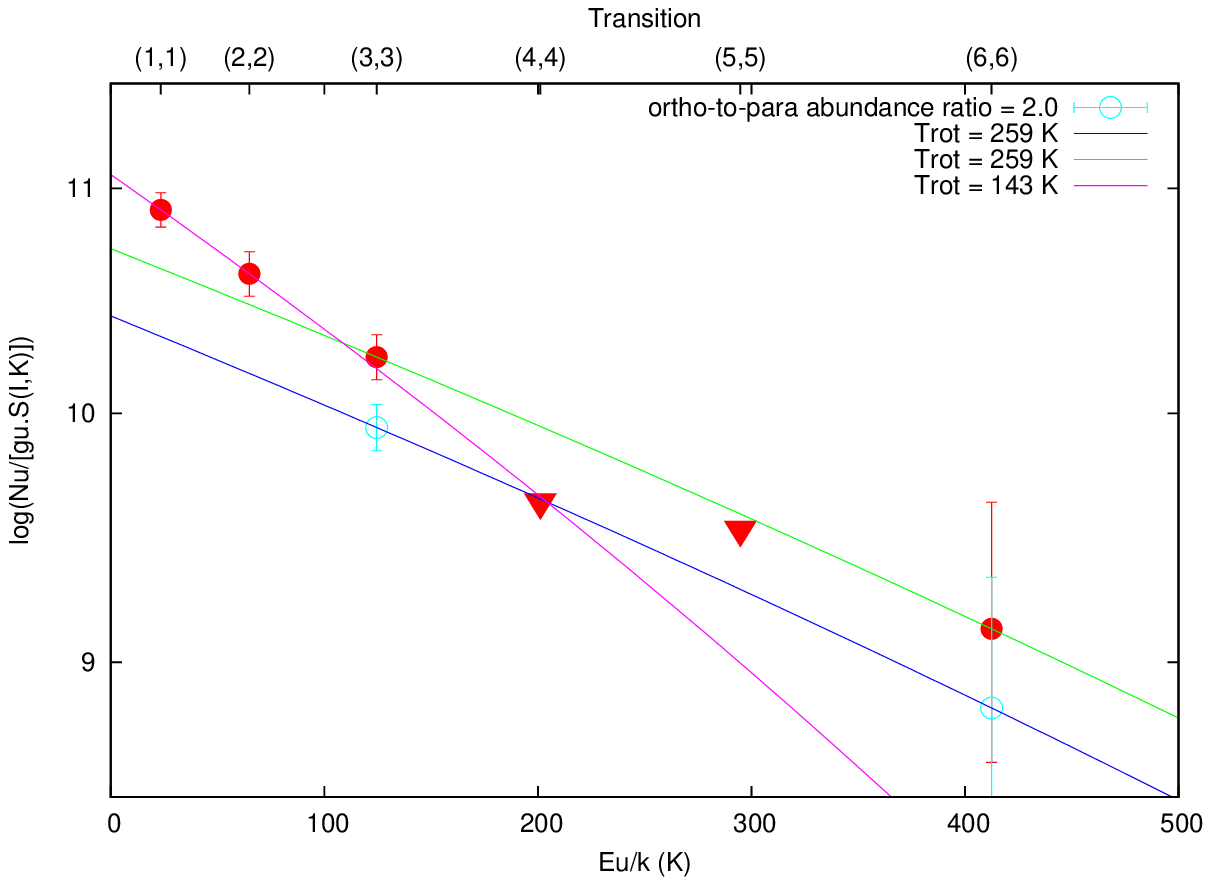}
\caption{As in Fig. \ref{1745_region1_Nuplot}, for R2 of HESS\,J1801$-$233 (shown in Fig. \ref{W28_33_to_11_ratio_image}). For this region an OPR of 1.5 was estimated. A lower limit OPR has been found (due to upper limits of NH$_3$\,(4,4) and (5,5) emission). This method indicates an OPR of > 2.0. As can be seen in this image however, the NH$_3$\,(3,3) and upper limit NH$_3$\,(4,4) column densities, with an OPR of 1.0, lie on the rotational temperature fit to the NH$_3$\,(1,1) and (2,2) emission (magenta line) indicating that the gas traced by these transitions may have an OPR of 1.0 and have a rotational temperature of that given by the NH$_3$\,(1,1) and (2,2) transitions. The NH$_3$\,(5,5) and (6,6) emission would then be tracing higher temperature gas, and the derived OPR would not reflect the conditions of this gas. This region is the pre shocked region of molecular gas associated with HESS\,J1801$-$233.}
\label{W28_region2_Nuplot}
\end{figure}
  
\subsection{TeV emission from PWNe} \label{PWNe}

Pulsar Wind Nebulae or old relic PWNe are thought to be responsible for the TeV emission from many of the unidentified Galactic gamma-ray sources \citep{2008ApJ...682.1177C,2009ApJ...694...12M}. The population of likely TeV PWNe is thought to include 27 sources \citep{2013APh....43..287D}, although many of these sources do not have associated, observed pulsars. \citet{2008AIPC..983..195G} have identified seven TeV sources as confirmed PWNe. We have added HESS\,J1303$-$631 to this list as it has been since confirmed as a PWN \citep{2012A&A...548A..46H}, and meets the same requirements for confirmed PWNe proposed by \citet{2008AIPC..983..195G}.

In order to assess whether TeV sources in our study meet energetics constraints provided by confirmed PWNe, we have assumed that the TeV emission is at the same distance as the dense molecular gas traced by NH$_3$\,(1,1) emission and scaled the TeV luminosity of each source accordingly. We then calculated the range of spin down powers a pulsar producing this emission could have if the TeV emission was between 0.01\% and 7\% of the pulsar spin down power (as is the case for the population of seven confirmed PWN in \cite{2008AIPC..983..195G}). In Fig. \ref{image:PSR_plot}, the range of TeV luminosities for all TeV sources included in this study, along with the range of spin down powers these luminosities would translate to, with minimum and maximum TeV efficiencies of 0.001\% and 7\% respectively. It can be seen that, even with minimum TeV efficiency, the range of inferred spin down powers is between 10$^{34}$erg/s and 10$^{39}$erg/s, a range consistent with the population of pulsars likely to have an association with TeV emission identified by \citet{2013APh....43..287D}. Therefore all of the TeV sources included in this study of dense gas detected by NH$_3$\,(1,1) adjacent to TeV emission meet energetics requirements to be Galactic pulsar wind nebulae.

In addition, we believe the range of inferred spin down powers in Fig. \ref{image:PSR_plot} indicates that there could be a population of older TeV emitting PWNe, in the population of TeV sources included in this study, with a higher TeV efficiency (i.e. close to the maximum current TeV efficiency seen of 7\%, or even greater), which do not yet have associated, detected pulsars considered likely to have an association with TeV emission identified by \citet{2013APh....43..287D}. This population of TeV emitting PWNe is likely to become more apparent with the increasing sensitivity of TeV gamma-ray observations.

This method has not allowed for discrimination based on energetics requirements for those sources which are currently unidentified, which may be PWNe. Fig. \ref{image:PSR_plot} demonstrates this as TeV luminosities derived for all molecular cores detected in our study fall within the expected range for TeV emitting PWNe. This study is useful for more detailed studies for individual TeV sources which compare the morphology of dense gas with the TeV emission \citep[e.g.][]{2016MNRAS.458.2813V}. In the case of PWNe, which are produced leptonically through inverse Compton emission, the TeV emission often exhibits asymmetric morphology, and is expected to anti-correlate with molecular gas. This asymmetry can be explained by an inhomogeneous distribution of interstellar gas around pulsars which strongly influences their development \citep[e.g.][]{2001ApJ...563..806B}. This study can be used to search for dense gas anti-correlated with asymmetric morphology of TeV emission to identify relic PWNe. In addition, these relic PWNe would be less likely to display 2-10 keV X-ray emission than the younger PWNe, in accordance with the ratio of gamma-ray to X-ray flux identified by \citet{2009ApJ...694...12M} to increase with PWNe age.

\begin{figure}
\centering
\includegraphics[width=0.5\textwidth, angle=0]{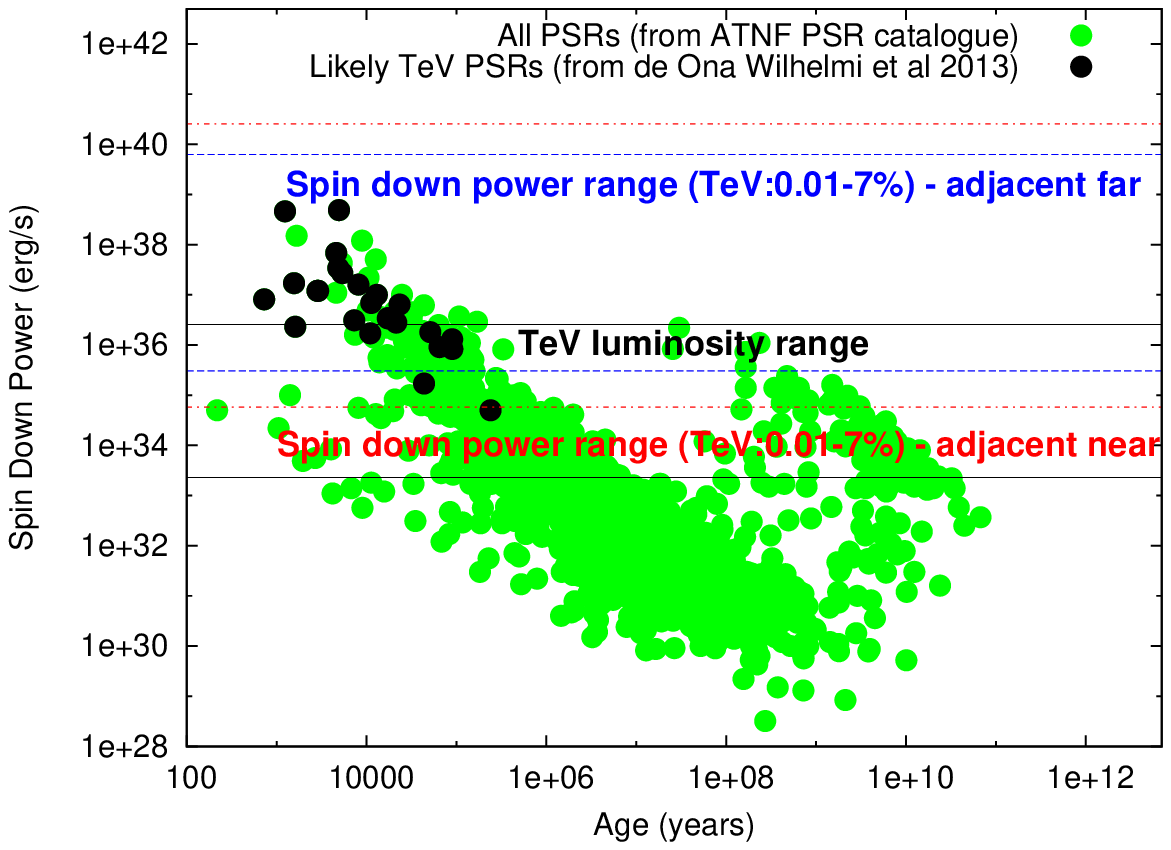}
\caption{Spin down power vs age for pulsars from the ATNF Pulsar Catalog \citep{2005AJ....129.1993M} plotted with green dots. The population of pulsars deemed as likely TeV PWNe by \citet{2013APh....43..287D} have been over-plotted with black dots. It can be seen that the minimum spin down power for those pulsars likely associated with TeV emission is $\sim 10^{34}$\,erg/s. The TeV luminosity range (scaled according to molecular clump distance) is indicated by black, solid lines. The range of spin down powers for minimum apparent TeV efficiencies of 0.001\% and maximum apparent TeV efficiencies of 7\% for adjacent clumps' near and far distances are indicated by blue dashed and red dot-dashed lines respectively.}
\label{image:PSR_plot}
\end{figure}


\section{Discussion of Six Prominent Unidentified TeV sources}\label{six_sources}

Six TeV sources which have not been unambiguously connected to any counterpart at other wavelengths are discussed in detail here. These sources were chosen for their coincidence with star formation regions as evident by IR emission, H$_2$O masers and/or H69$\alpha$ emission or, as for HESS\,J1745-303, an anomalous NH$_3$ ortho-to-para brightness temperature ratio. Images of the peak pixel brightness temperature between velocities of $-200$km/s and $200$km/s of various molecular transitions along with spectra towards selected molecular gas clumps and infra-red features are displayed in Figures \ref{J1745m303_image} to \ref{J1729m345_IR_image}. A discussion of notable gas and infra-red features is included in the following section. The molecular transition gas maps seen in Fig. \ref{J1745m303_image} to \ref{J1729_J1731_image} have been chosen based on significant emission \textbf{(defined as > 3$\sigma$, where $\sigma \equiv {\rm{T}_{\rm{rms}}}/{\sqrt{5}}$)} seen towards or adjacent to the TeV emission. Maps of the other transitions searched for are not shown. Kinematic distance solutions and molecular clump masses for the dense gas (traced by NH$_3$ (1,1) emission) towards, and adjacent to, the six TeV sources discussed in this section are presented in Table \ref{table:six_sources_mass_etc}.

For the sources for which the TeV could possibly originate from cosmic ray protons, a simple calculation of diffusion time has been used to help assess the plausibility of the TeV emission coming from interactions between cosmic rays from a nearby accelerator and dense gas catalogued in this survey.
 
To calculate the distance $d$ travelled by cosmic ray protons in a given time, $t$, we have used:
\begin{equation}
  d=\sqrt{2D(E_p,B)t}
\label{equation:diff_dist}
\end{equation}
where $D(E_p,B)$ is the diffusion coefficient dependent on maximum proton energy, $E_p$ and magnetic field $B$ according to:
\begin{equation}
D(E_p,B) = \chi D_0 (\frac{\rm{E_p/GeV}}{B/3\mu\rm{G}})^{0.5}\,\rm{cm}^2\,\rm{s}^{-1}
\label{equation:diff_coeff}
\end{equation}
from \citet{2007Ap&SS.309..365G}. $D_0$ is the average Galactic diffusion coefficient \citep{1990acr..book.....B}, $\chi$ is the diffusion suppression coefficient 
(assumed to be $<1$ inside molecular cores; 1 outside) \citep{1990acr..book.....B,2007Ap&SS.309..365G}. The magnetic field strength as a function of density $n_{H_2}$\,cm$^{-3}$ using: 
\begin{equation}
  B(n_{H_2}) \sim 10(\frac{n_{H_2}}{300\mu\rm{G}})^{0.65} \mu\rm{G} 
\label{equation:Crutcher_Bfield}
\end{equation}
was taken from \citet{1999ApJ...520..706C} based on their Zeeman splitting measurements in molecular clouds.

\begin{table*}
\caption{All molecular detections from molecular clumps with NH$_3$\,(1,1) emission towards Galactic TeV sources. This table shows an extract from the full table, which is available from the online appendix. For individual sources such as HESS\,J1745$-$303, the large spread of distances between different molecular clouds, may be due to the clouds being located at different distances along the line of sight, or perhaps due to intrinsic movement of the molecular clouds. HESS\,J1745$-$303 is located close to the CMZ, and so the Galactic rotation curve used here does not include solutions for the distances to several molecular clouds observed. For HESS\,J1745$-$303 we believe the clouds are separated into several distinct distance groups. These distance ambiguities will investigated further in an upcoming paper (de Wilt, P. \& Rowell, G. in preparation).}
\centering
\begin{tabular}{|c c c c c c c c c|}
\hline
TeV Source			&	NH$_3$	&	NH$_3$	&	NH$_3$\,(1,1)				& kinematic & distance	&	H$_2$ 	&$N_{NH_3}$	&	H$_2$	\\
					&	(1,1)	&	(1,1)	&\textit{V$_{\rm{LSR}}$}	&near		&far		&mass (near)& $\times10^{13}$&	density \\
					&\textit{l}	&\textit{b}	&	 (km/s)					&(kpc)		&(kpc)		&M$_{\odot}$	&(cm$^{-2}$)&(cm$^{-3}$)\\\hline
HESS\,J1626$-$490	&	334.71	&	0.04	&	$-$86.46				&	5.5		&	 8.6	&	 0.2	&	7.7		&1.3$\times10^3$	\\
HESS\,J1640$-$465	&	338.09	&	0.01	&	$-$40.00				&	3.3	  	&	11.2	&	1153	&	8.4		&5.8$\times10^2$	\\
					&	338.33	&	0.14	&	$-$36.59				&	3.1		&	11.4	&	 100	&	1.3		&1.2$\times10^2$	\\
					&	338.47	&	0.04	&	$-$37.01				&	3.2		&	11.3	&    975	&	5.2		&3.1$\times10^2$	\\
HESS\,J1729$-$345	&	353.28	&	$-$0.06	&	$-$16.09				&	3.9		&	10.2	&	 -		&	-		& -			\\
HESS\,J1745$-$303	&	358.37	&	$-$0.46	&	    6.50				& 	-		& 	-		&	353		&	1.2		&0.6$\times10^2$	\\
					&	358.46	&	$-$0.38	&	 $-$3.73				&	3.3		&	9.3		&	186		&	2.4		&2.2$\times10^2$	\\
					&	358.59	&	$-$0.81	&	    2.66				&	19.4	&  	-		&	-		&			&			\\
					&	358.64	&	$-$0.40	&	 $-$6.71				&	6.3		& 	-		&	0.1		&	0.4		&6.7$\times10^2$	\\
					&	358.80	&	$-$0.36	&	$-$29.73				& 	-		& 	-		&	23827	&	11.1	&2.0$\times10^2$	\\
					&	358.81	&	$-$0.51	&	$-$54.02				& 	-		& 	-		&	-		&			&	\\	
HESS\,J1804$-$216	&	8.14	&	   0.22	&	   17.96				&	3.7		&	10.5	&	517		&	3.0		&1.8$\times10^2$	\\
					&	8.25	&	   0.17	&	   17.10				&	3.5		&	10.7	&	212		&	1.4		&1.0$\times10^2$	\\
					&	8.40	&	$-$0.28	&	   36.71				&	6.7		&  	-		&	7669	&	3.4		&0.6$\times10^2$	\\
					&	8.68	&	$-$0.40	&	   36.71				&	6.4		& 	- 		&	8091	&	4.8		&1.0$\times10^2$ \\
HESS\,J1848$-$018	&	30.72	&	$-$0.07	&	   92.16				&	5.2		&	9.2		&	1942	&	6.3		&2.9$\times10^2$ \\
					&	30.78	&	$-$0.08	&	   94.30				&	5.1		&	9.1		&	1.2		&	4.8		&8.1$\times10^3$ \\
					&	30.81	&	$-$0.04	&	   93.87				&	5.1		&  	9.1		&	1909	&	5.9		&2.6$\times10^2$ \\
					&	30.98	&	$-$0.14	&	   77.25				&	4.4		& 	10.0 	&	-		&	-		&	-		\\\hline
					
\end{tabular}
\label{table:six_sources_mass_etc}
\end{table*}

\subsection{HESS\,J1745$-$303}

\textbf{HESS\,J1745$-$303} in Fig. \ref{J1745m303_image} is a source where NH$_3$\,(1,1) emission is coincident with the TeV peak. Of particular interest here is the broad line emission of all thermal molecular line transitions towards the TeV centroid. HESS\,J1745$-$303 partially overlaps both the molecular cloud and the SNR G$359.1-0.5$. The SNR G$359.1-0.5$ interacts with a molecular cloud as evident by a cluster of 1720 MHz OH masers towards the SNR rim. This SNR has been attributed to producing this broad line emission, however the broad line emission has a kinematic velocity which differs from that of the masers by around 50 km/s and the broad emission line region lies outside the boundary of the SNR observed in radio continuum (see Fig. \ref{J1745m303_image} where the SNR boundary is indicated). Broad line molecular gas is only observed towards one region of the gamma-ray emission, defined as region A by \citet{2008A&A...483..509A} (and indicated in Fig. \ref{J1745m303_image}), which includes the highest TeV gamma-ray significance region, the 7$\sigma$ contour. The molecular line emission detected in our study, within the 7$\sigma$ significance contour, includes this broad emission line cloud (FWHM > 10\,km/s), the spectra of which are displayed in the LHS of Fig. \ref{J1745m303_image}. In addition to broad emission lines, NH$_3$\,(3,3) emission is observed to have a higher peak brightness temperature than NH$_3$\,(2,2) and (1,1). As discussed in Section \ref{ortho-para}, this molecular cloud has an enhanced ortho-to-para NH$_3$ abundance ratio which we believe is due to the previous passage of a shock through the cloud. The nature of the shock is not clear as this gas is outside the boundary of the observed SNRs in the region. Non-thermal broadening (i.e. from turbulence) of these spectral lines dominates over the thermal broadening ($<$ 4 km/s) which further supports that a shock as passed through this molecular cloud. This will be discussed in more detail in a future study (de Wilt, P. \& Rowell, G. in preparation).

There are several other potential sources of cosmic-ray hadrons and/or electrons in the region which include two energetic pulsars, an additional SNR, G359.0-0.9, (all of which are indicated in Fig. \ref{J1745m303_image}) and a further radio continuum feature seen in archival data taken as part of the Molonglo Galactic Plane Surveys \citep{1999ASPC..168...43G}, which, could potentially be another SNR (de Wilt, P. \& Rowell, G. in preparation).

Previous studies have discussed the lack of molecular cloud emission towards the West and South-West components of TeV emission in HESS\,J1745$-$303 \citep[e.g.][]{2008A&A...483..509A, 2012PASJ...64....8H}. Our study confirms a lack of broad molecular line emission towards these components of TeV emission, and therefore suggests that there is no Galactic ridge molecular cloud counterparts to the West and South-West components of the TeV source. We have identified dense molecular gas through NH$_3$ transitions which are found towards region B (as defined by \citet{2008A&A...483..509A} and seen in Fig. \ref{J1745m303_image}) which may be associated with the TeV emission. The NH$_3$ emission from these dense gas cloud cores exhibits narrow linewidths ($<$\,4\,km/s) which suggests that the emission is foreground or background to the Galactic Centre. \citet{2012MNRAS.426.1972P} state that the sensitivity of the HOPS observations mean that it is unlikely that any emission seen is background to the Galactic Centre, which makes it likely that these clouds are foreground to the GC. The association between this newly detected gas and the TeV emission is being investigated further and the results will be presented in a future paper. The narrow line NH$_3$\,(1,1) emission extent and spectrum can be seen in Fig. \ref{J1745m303_image}.

Fig. \ref{J1745m303_IR_image} shows that no active star formation is seen overlapping the TeV centroid in either the Spitzer infra-red bands nor in the molecular tracers used in our study. A similar situation is seen towards HESS\,J1801$-$233, the TeV peak towards the interacting SNR W28 \citep{2011MNRAS.411.1367N}. The lack of IR emission here supports the interpretation of the broad line emission not being associated with heating due to star formation processes, but perhaps due to additional energetics such kinetic energy provided by a shock which has passed through the gas cloud. As the gas lies outside the boundary of the SNR G359.1-0.5 as seen in radio continuum, we propose that there has been another shock which has passed through the cloud, producing the post-shocked gas characteristics of broad line emission, an ortho-to-para brightness temperature ratio > 1 as well as extended SiO\,(1-0) emission.

\begin{figure*}
\centering
\includegraphics[width=0.95\textwidth, angle=0]{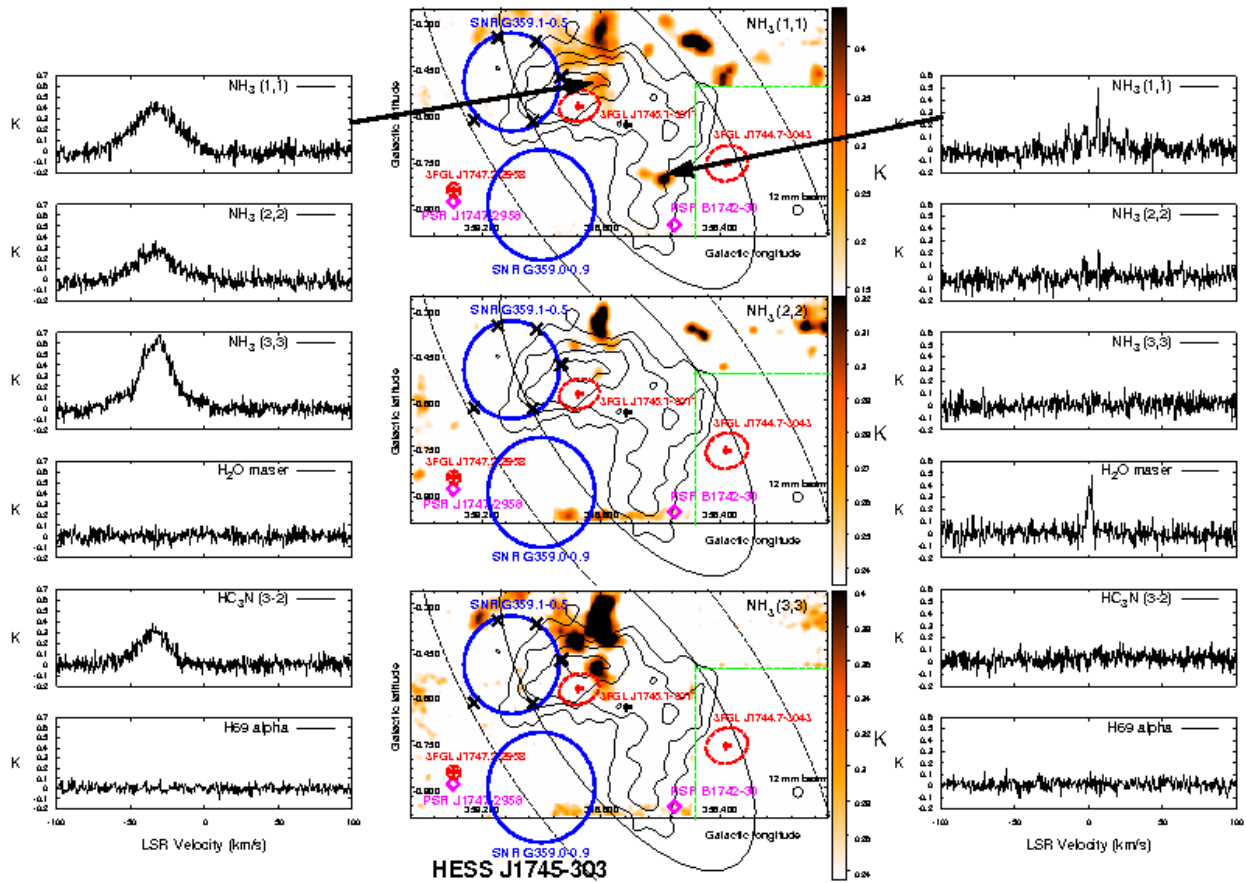}
\caption{12\,mm peak pixel maps of molecular line emission towards the TeV source \textbf{HESS\,J1745$-$303}. TeV significance contours (4-7$\sigma$) \citep{2008A&A...483..509A} are shown in black. Images of the peak pixel along the line of sight between local standard of rest (LSR) velocities $-$200 and 200 km/s of NH$_3\,(J,K), (J=K=1,2,3)$ are seen. Broad line emission is displayed in the spectra of all thermal lines seen towards the 7$\sigma$ TeV significance contour, however no H$_2$O masers or H69$\alpha$ emission is detected in our study. Spectra from a dense cloud core towards the SW component of TeV emission is seen (on the RHS here) offset in velocity from the broad emission by $\sim$55 km/s. The size of the region each of the spectra displayed in this image and Figs \ref{J1640m465_image}, \ref{J1848m018_image}, \ref{J1626m490_image}, \ref{J1804m216_image} and \ref{J1729_J1731_image} is taken from is given by one 12\,mm beam size (displayed in each image).}
\label{J1745m303_image}
\end{figure*}

\begin{figure*}
\includegraphics[width=0.75\textwidth]{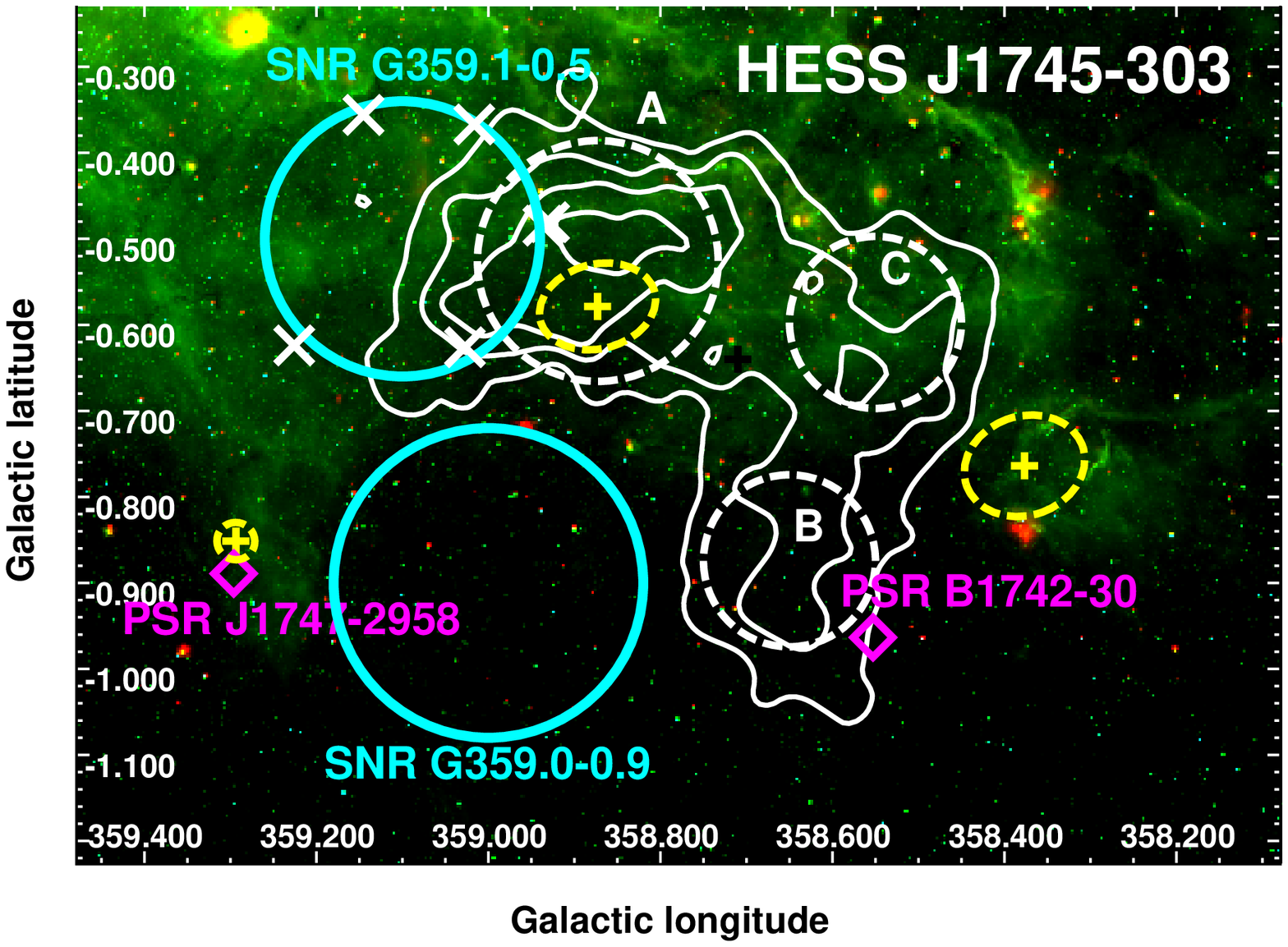}
\caption{\textit{Spitzer GLIMPSE/MIPSGAL} three-colour (RGB=24/8/4.5$\mu m$)M Jy sr$^{-1}$) image of the region towards HESS\,J1745$-$303 (the TeV emission is indicated by white contours). Several 'rings' can be seen in the green, 8$\mu m$ emission, however there is no widespread star formation shown in these IR bands towards the TeV emission.}
\label{J1745m303_IR_image} 
\end{figure*}

\subsection{HESS\,J1640$-$465}
Fig. \ref{J1640m465_image} displays NH$_3$\,(1,1) emission towards the edges of the TeV gamma-ray source \textbf{HESS\,J1640$-$465}. The gamma-ray emission is seen towards the SNR G338.3$-$0.0 and a giant H{\sc ii} region G338.4+0.1, both detected in radio continuum at 843 MHz \citep{1996A&AS..118..329W}. The faint ASCA X-ray source AX J1640.7$-$4632 \citep[see][]{2001ApJS..134...77S} is coincident with HESS\,J1640$-$465, as is the Fermi-LAT source 3FGL J1640.4$-$4634. A separate TeV source named HESS\,J1641$-$463, with a harder spectral index, to the edge of this TeV source is seen \citep[see][]{2014ApJ...794L..16L} towards the SNR G338.5+00.1. 
The molecular line emission seen in HOPS data shows broad ($\sim 5$\,km/s) NH$_3$ emission in the (1,1) to (3,3) inversion transitions, indicative of shocked/excited gas towards the giant H{\sc ii} region G338.4$+$0.1, located between the two TeV sources. Several H$_2$O masers \citep{2011MNRAS.416.1764W} are seen towards and surrounding this TeV source (seen in Fig. \ref{J1640m465_image}), indicating the ongoing star formation in the region. All of the detected molecular line transitions have a far kinematic distance matching that of both of the SNRs in the region ($\sim$11 kpc, \citealt{2007A&A...468..993K}).

The pulsar PSR J1640$-$4631 has been recently discovered by \citet{2014ApJ...788..155G}, using the Nuclear Spectroscopic Telescope Array (NuSTAR), towards HESS\,J1640$-$465 with a a spin-down luminosity of 4.4$\times10^{36}$ erg s$^{-1}$. The pulsed X-ray emission and the TeV centroid of HESS\,J1640$-$465 occur at the centre of the SNR G338.3$-$0.1. This, together with the morphology of the molecular material detected in our study (which surrounds the TeV emission), may favour a PWN model for the TeV emission. Rescaling the TeV flux (> 1\,TeV) to the distance of the SNRs and NH$_3$ emission gives a gamma-ray luminosity of $2.4\times10^{34}$\,erg/s. Assuming that PSR\,J1640$-$4631 is linked to SNR G338.2$-$0.0, and so is at a distance of 11 kpc, and is powering the PWN HESS\,J1640$-$465, the apparent efficiency of PSR\,J1640$-$4631 would be $\sim$0.05\%, within the range of confirmed TeV PWNe, $0.01$ to $7\%$ \citep{2008AIPC..983..195G}.

3FGL\,J1640.4$-$4634, also catalogued as 1FGL\,J1640.8$-$4634, the Fermi-LAT GeV source coincident with HESS\,J1640$-$465, has a spectrum (from five years of Fermi-LAT data) which joins smoothly to the TeV spectrum \citep{2014ApJ...794L..16L}. This smooth spectrum from GeV to TeV favours a hadronic origin to the TeV gamma-rays, and has been reproduced by a hadronic model of HESS\,J1640$-$465 \citep{2014ApJ...794L..16L}, however the spectrum can also be reproduced by fine tuning to a leptonic, PWN model \citep[see][]{2014ApJ...788..155G}, and so this possibility cannot be ruled out.

Adjacent to HESS\,J1640$-$465 is the TeV source HESS\,J1641$-$463 \citep{2014ApJ...794L...1A}. This source has a harder spectral index than HESS\,J1640$-$465 and several possible counterparts including SNR G338.5$+$0.1, several X-ray sources (which may point towards a PWN scenario) have been identified \citep{2014ApJ...794L...1A}. HESS\,J1641$-$463 could also be a TeV binary system, although no variability in TeV gamma-rays or X-rays has been observed in current data \citep{2014ApJ...794L...1A}. The spectrum of the GeV source 3FGL\,J1641.1$-$4619c is very soft \citep[see][]{2014ApJ...794L..16L}, compared to the hard TeV spectrum detected by H.E.S.S. \citep{2014ApJ...794L...1A} which suggests different origins for the GeV and TeV gamma-rays. \citet{2017MNRAS.464.3757L} have used observations of many molecular and atomic transitions towards HESS\,J1640$-$465 and HESS\,J1641$-$463 to gain an understanding of total gas distribution and mass to explore both hadronic and leptonic TeV emission scenarios.
\citet{2017MNRAS.464.3757L} find that in a hadronic origin for the gamma-ray emission, the cosmic-ray enhancement rates are $\sim10^3$ and $\sim10^2$ times the local solar value for HESS\,J1640$-$465 and HESS\,J1641$-$463 respectively. In the hadronic case, if HESS\,J1641$-$465 were produced by runaway protons from the SNR G338.3$-$0.0 and applying the diffusion distance estimate according to Equation \ref{equation:diff_dist}, protons of 100 TeV would take $\sim$10,000 yrs to travel from the SNR to the second TeV source if moving through the dense gas. In this case, however, we would expect the TeV gamma-rays to peak along with the density of molecular gas, which, as can be seen in Fig. \ref{J1640m465_image} appears not to be the case. As outlined by \citet{2017MNRAS.464.3757L}, and which can be seen in Table \ref{table:all_sources}, in the highest density region, there are molecular cloud cores at different distances along the line of sight, which could allow for CR diffusion between them. This would allow for the CRs to diffuse faster, and using the diffuse gas parameters of \citet{2017MNRAS.464.3757L}, we calculate that 100 TeV protons would take $\sim$1600 yrs to diffuse from SNR G338.3$-$0.0 to HESS\,J1641$-$463. As the age of SNR G338.3$-$0.0 has been estimated from 1-2 kyr up to 5-8 kyr, this would allow time for VHE CRs to travel from SNR G338.3$-$0.0 to produce the TeV emission from HESS\,J1641$-$463, between dense cloud cores (as outlined in \citet{2017MNRAS.464.3757L}).

\begin{figure*}
\centering
\includegraphics[width=0.95\textwidth, angle=0]{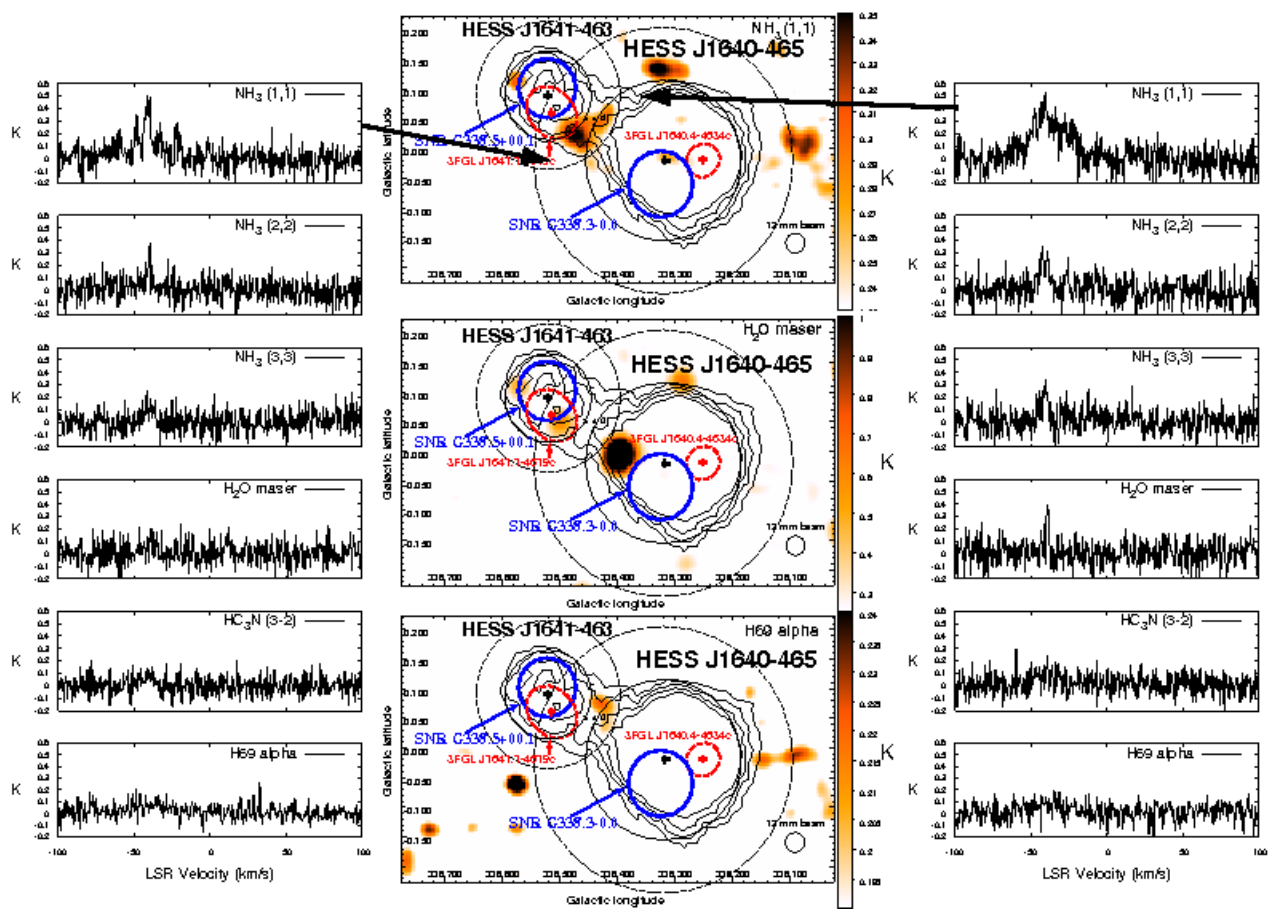}
\caption{12\,mm peak pixel maps of molecular line emission towards the TeV sources \textbf{HESS\,J1640$-$465} and \textbf{HESS\,J1641$-$463} (which are indicated by solid, black contours). Three gas clumps are seen traced by NH$_3$\,(1,1) emission, all adjacent to the TeV emission. Two of these clumps (for which the spectra are shown above) display H$_2$O maser emission and H69$\alpha$ emission indicating both ongoing star formation and ionised gas.}
\label{J1640m465_image}
\end{figure*}

\begin{figure*}
\includegraphics[width=0.75\textwidth]{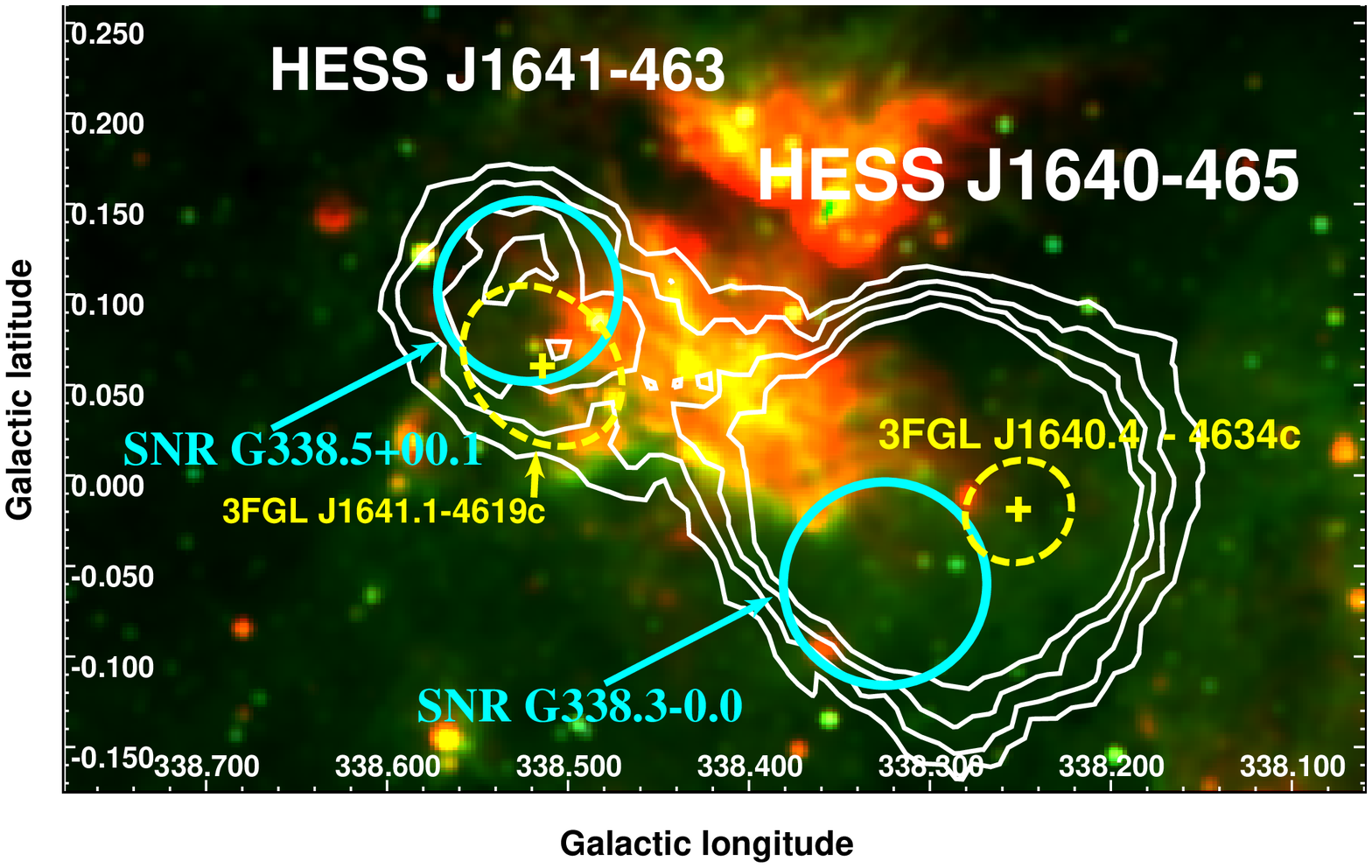}
\caption{\textit{Spitzer GLIMPSE/MIPSGAL} three-colour (RGB=24/8/4.5$\mu m$)M Jy sr$^{-1}$) image of the region towards HESS\,J1640$-$465 and HESS\,J1641$-$463. Two giant H{\sc ii} complexes are seen as prominent features of this image. G338.4$+$0.1 lies between the two TeV sources. These H{\sc ii} complexes are both traced by H69$\alpha$ in our study (seen in Fig. \ref{J1640m465_image}.}
\label{J1640m465_IR_image}
\end{figure*}

\subsection{HESS\,J1848$-$018}
\textbf{HESS\,J1848$-$018} is located towards the W43 star forming region. No obvious counterparts for the TeV emission such as SNRs or energetic pulsars are seen towards HESS\,J1848$-$018 \citep{2008AIPC.1085..372C}. The high mass stellar cluster W43 is offset from the centroid of the TeV emission within a giant molecular cloud with mass $\sim10^6$\,M$_\odot$ detected in both sub-millimetre wavelengths \citep{2003ApJ...582..277M} and $^{13}$CO \citep{2006ApJS..163..145J}. This cluster contains the high mass binary system, WR 121a,  catalogued by \cite{2001NewAR..45..135V} and classified as a colliding wind binary (CWB) by \citet{2011ApJ...727..105A}. Our observations (Fig. \ref{J1848m018_image}) reveal broad NH$_3$ emission lines in the giant molecular cloud as well as H$_2$O maser emission and H69$\alpha$ emission indicative of ongoing star formation and ionised gas towards WR\,121a. The H69$\alpha$ emission traces the H{\sc ii} region G30.8$-$0.0 \citep{1985ApJ...296..565L}.
The infra red dark cloud (IRDC) G030.97$-$00.14, catalogued by \citet{2006ApJ...641..389R}, is seen towards the TeV centroid in Fig. \ref{J1848_IR_image}, overlapping a gas clump detected in NH$_3$\,(1,1) by our study. Spectra towards this gas clump are displayed on the LHS of Fig. \ref{J1848m018_image}, which show H$_2$O maser emission, indicative of active star formation. Due to the coincident IRDC, we suggest that this is an early star formation region, slightly foreground to the W43 region which provides much of the extended, background 8$\mu$m emission in Fig. \ref{J1848_IR_image} which is being absorbed by the cold, dense gas.

Colliding wind binaries are expected to produce leptonic gamma-ray emission through Inverse Compton (IC) scattering up to a few GeV \citep[e.g.][]{2006MNRAS.372..801P,2013A&A...558A..28D}. Theoretical models suggest that, given a suitable environment, stellar winds which are ejected during the high mass star evolution may also be able to accelerate hadrons as effectively as SNRs \citep[see][and references therein]{2011A&A...526A..57F}. If the CWB WR 121a is indeed responsible for the TeV emission, then approximately 3$\%$ of the kinetic power of the wind-wind collision (estimated to be $\sim10^{36}$\,erg.s$^{-1}$ in the system WR140 \citep{2006MNRAS.372..801P}) is converted into TeV emission. This agrees with numerical simulations of relativistic collision-less shocks, which indicate that at least 10$\%$ of energy is converted into relativistic particles downstream of the shock \citep{2008ApJ...682L...5S}.

Coincident with HESS\,J1848$-$018 are the Fermi-LAT source 3FGL J1848.4$-$0141 \citep{2015ApJS..218...23A} and HAWC source 1HWCJ1849$-$017c \citep{2016ApJ...817....3A}. While spectra of the HESS and HAWC sources have not been published, a continuous spectrum from Fermi-LAT energies up to tens of TeV (where HAWC is most sensitive) would point towards a hadronic origin to the gamma-rays, as leptonic sources are expected to cut off at lower energies. Further studies to determine the TeV morphology of HESS\,J1848-018 would be beneficial in determining whether the CWB WR 121a is a possible counterpart to the TeV emission. The cold, dense core detected by NH$_3$ observations in our study towards the IRDC G030.97$-$00.14 is within 2 arcmin of the TeV centroid and could provide a target for accelerated hadrons. A study of the morphology of more diffuse gas (i.e. that traced by $^{12}$CO(1-0) and H{\sc i}) is planned, and will give clues as to the origin of the TeV gamma-rays.

\begin{figure*}
\centering
\includegraphics[width=0.95\textwidth, angle=0]{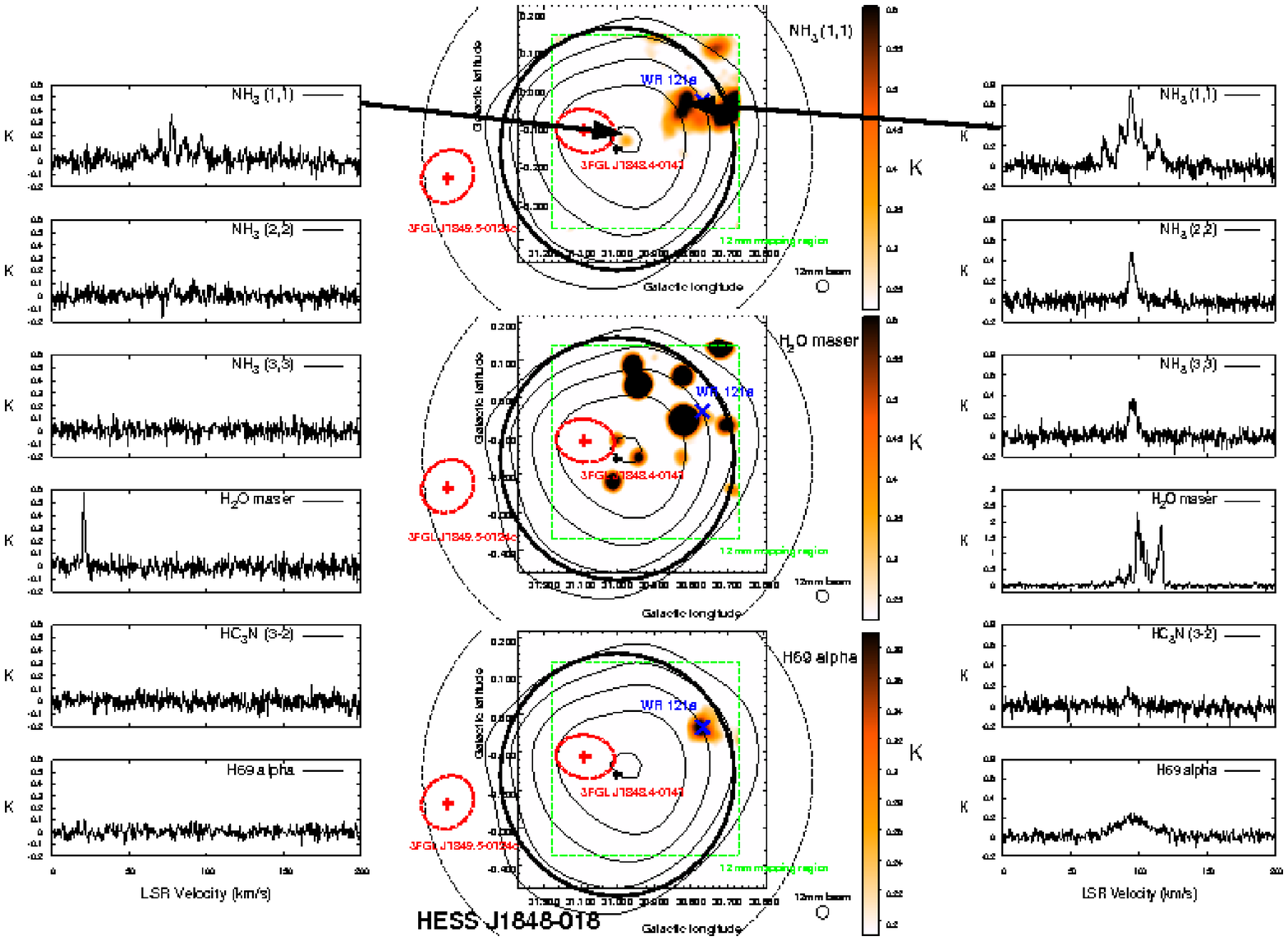}
\caption{12\,mm peak pixel maps of molecular line emission towards the TeV source \textbf{HESS\,J1848-018}. The centroid position of TeV emission is indicated by a + and intrinsic size by the dashed, circle. Peak brightness temperature images (as described in Fig. \ref{J1745m303_image}) of NH$_3$\,(1,1) , H$_2$O maser and H69$\alpha$ are shown. The mapping region is indicated by a green, dashed box and does not cover the extent of the TeV emission. The RHS spectra display emission in all molecular line transitions included in our study, indicating dense gas along with ongoing star formation and ionised gas towards the stellar cluster W43. The LHS spectra display emission in NH$_3$\,(1,1) and NH$_3$\,(2,2) indicating dense gas and H$_2$O emission indicating ongoing star formation towards the TeV emission centroid.}
\label{J1848m018_image}
\end{figure*}

\begin{figure*}
\includegraphics[width=0.75\textwidth]{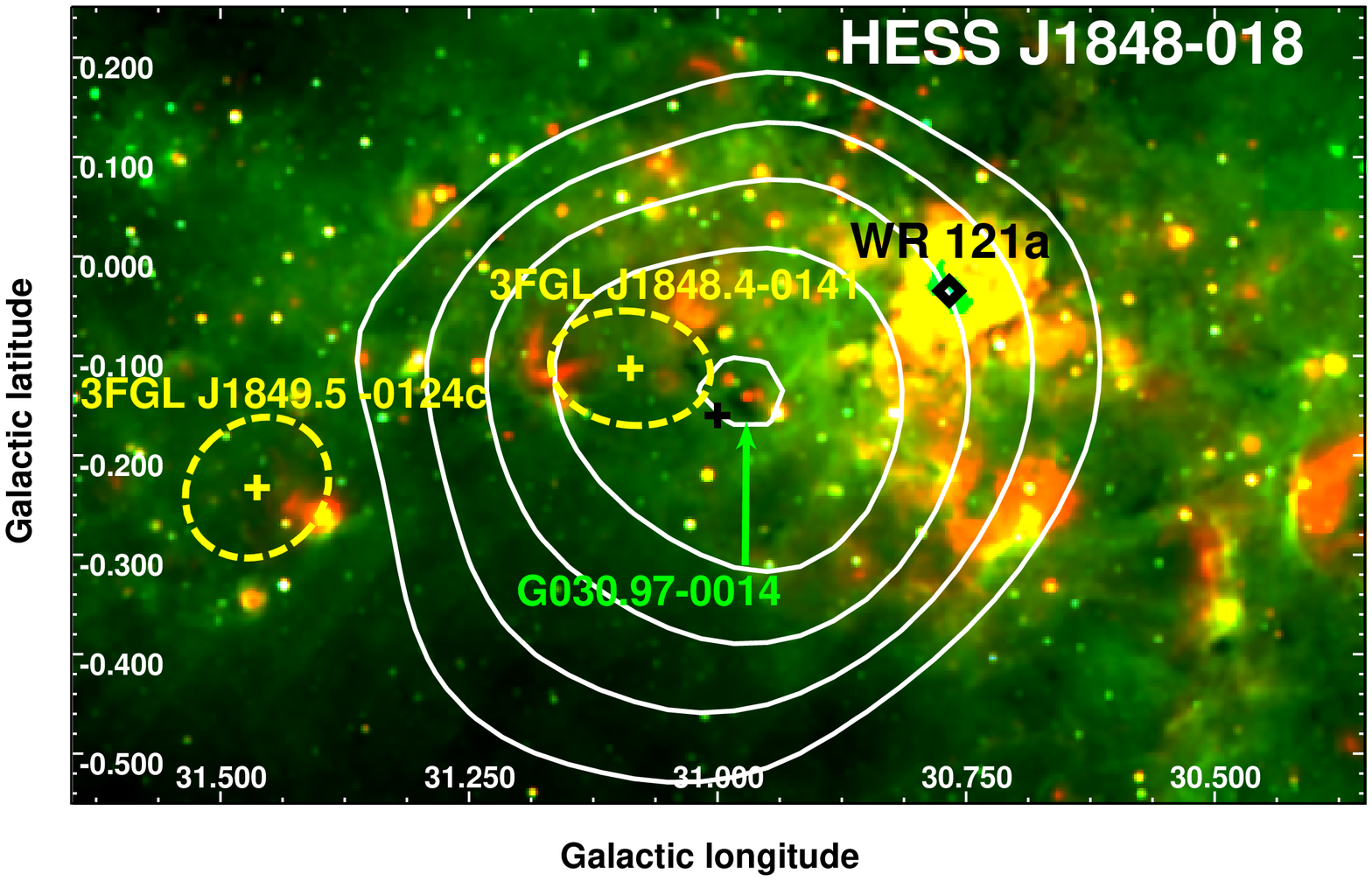}
\caption{\textit{Spitzer GLIMPSE/MIPSGAL} three-colour (RGB=24/8/4.5$\mu m$)M Jy sr$^{-1}$) image of the region towards HESS\,J1848$-$018. The TeV emission is displayed as in \ref{J1848m018_image}. Notable features in this IR image include a saturation of 24$\mu m$ emission towards the W43 stellar cluster which contains CWB WR 121a as well as excess emission seen in all IR bands surrounding this region. The IRDC G030.97-00.14 is also seen, towards the TeV centroid (indicated by a black cross), which matches the position of the gas clump traced by NH$_3$\,(1,1) emission seen in Fig. \ref{J1848m018_image}.}
\label{J1848_IR_image}
\end{figure*}

\subsection{HESS\,J1626$-$490}
The centroid of \textbf{HESS\,J1626$-$490} is found within 0.5$\degree$ of the SNR G$335.2+0.1$ which has been noted as a possible source of accelerated particles responsible for the TeV emission \citep[see][]{2011A&A...526A..82E}. An H{\sc i} void is seen in Southern Galactic Plane Survey \citep[SGPS,][]{2005ApJS..158..178M} data between velocities of $-24$ and $9$ km/s which is thought to correspond to the SNR \citep{2011A&A...526A..82E}. The GeV Fermi-LAT source 3FGL\,J1626.2$-$4911 is coincident with HESS\,J1626$-$490, however the GeV and TeV spectra are not well matched and so the sources may not be associated.
Our study reveals NH$_3$\,(1,1) emission towards the TeV peak which can be seen in Fig. \ref{J1626m490_image}. The NH$_3$\,(1,1) emission has a kinematic velocity of -86km/s, giving implied kinematic distance solutions $\sim$5.5 kpc (near) and $\sim$8.6 kpc (far). The position of the molecular clump traced by NH$_3$ emission in our study corresponds to that of an IRDC which can be seen in Fig. \ref{J1626m490_IR_image}. This indicates that it is foreground to the widespread IR emission. This IRDC, MSXDC G334.70+0.02, has been previously catalogued by \citet{2008ApJ...680..349J}. The widespread IR emission is assumed to originate from the complex containing the H{\sc ii} region G$334.684-0.107$, which has a distance of 12.8 kpc \cite{2005A&A...429..497R}. Both the near and far kinematic distances of the NH$_3$ clump are foreground to the H{\sc ii} region so this does not resolve the distance ambiguity.
HI absorption towards the SNR G335.2+0.1 seen in HI data from SGPS \citep{2005ApJS..158..178M} indicates a SNR distance between 5.9 and 12.4 kpc. Assuming the SNR and NH$_3$ detected molecular core are at the same distance, we can assign the far distance of 8.7 kpc to the NH$_3$ (1,1) emission detected in this study. At a distance of 8.7 kpc, the distance for CR protons to travel from the SNR shell to the far side of HESS\,J1626-490 is 76 pc. Using equations \ref{equation:Crutcher_Bfield} and \ref{equation:diff_dist} and assuming an average gas density of 1$\times10^2\,$cm$^{-3}$ towards the TeV emission, $10^{13}\,$eV CR protons would take $\sim$5,300\,yrs to traverse this distance, and lower energy cosmic rays would take a longer time. This would indicate a middle-aged or old SNR, at an age similar to other TeV emitting SNRs where a hadronic scenario is supported.

\begin{figure*}
\centering
\includegraphics[width=0.95\textwidth, angle=0]{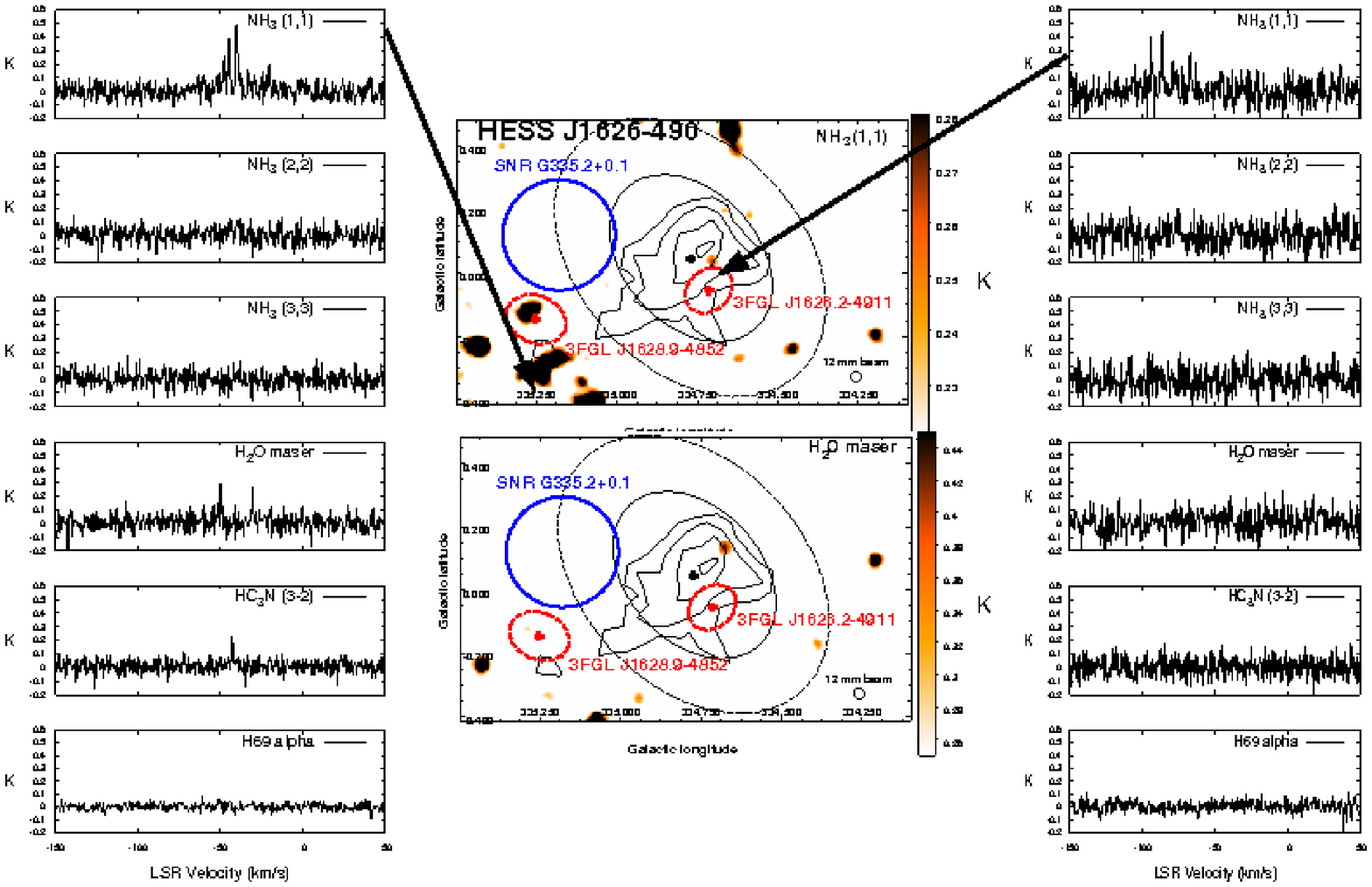}
\caption{12\,mm peak pixel maps of molecular line emission towards the TeV source \textbf{HESS\,J1626$-$490}. TeV emission is indicted by black contours, Fermi-LAT sources are indicted by dashed, red ellipses, the SNR G335.2$+$0.1 is indicated by a blue circle. Very little molecular emission towards this TeV source is detected in our study. A dense clump, traced by NH$_3$\,(1,1) emission is seen towards the TeV peak which also displays H$_2$O maser emission. Several molecular clumps are seen, traced by emission from several transitions, are seen below the SNR G335.2$+$0.1 in this image, away from the TeV emission.}
\label{J1626m490_image}
\end{figure*}

\begin{figure*}
\includegraphics[width=0.75\textwidth]{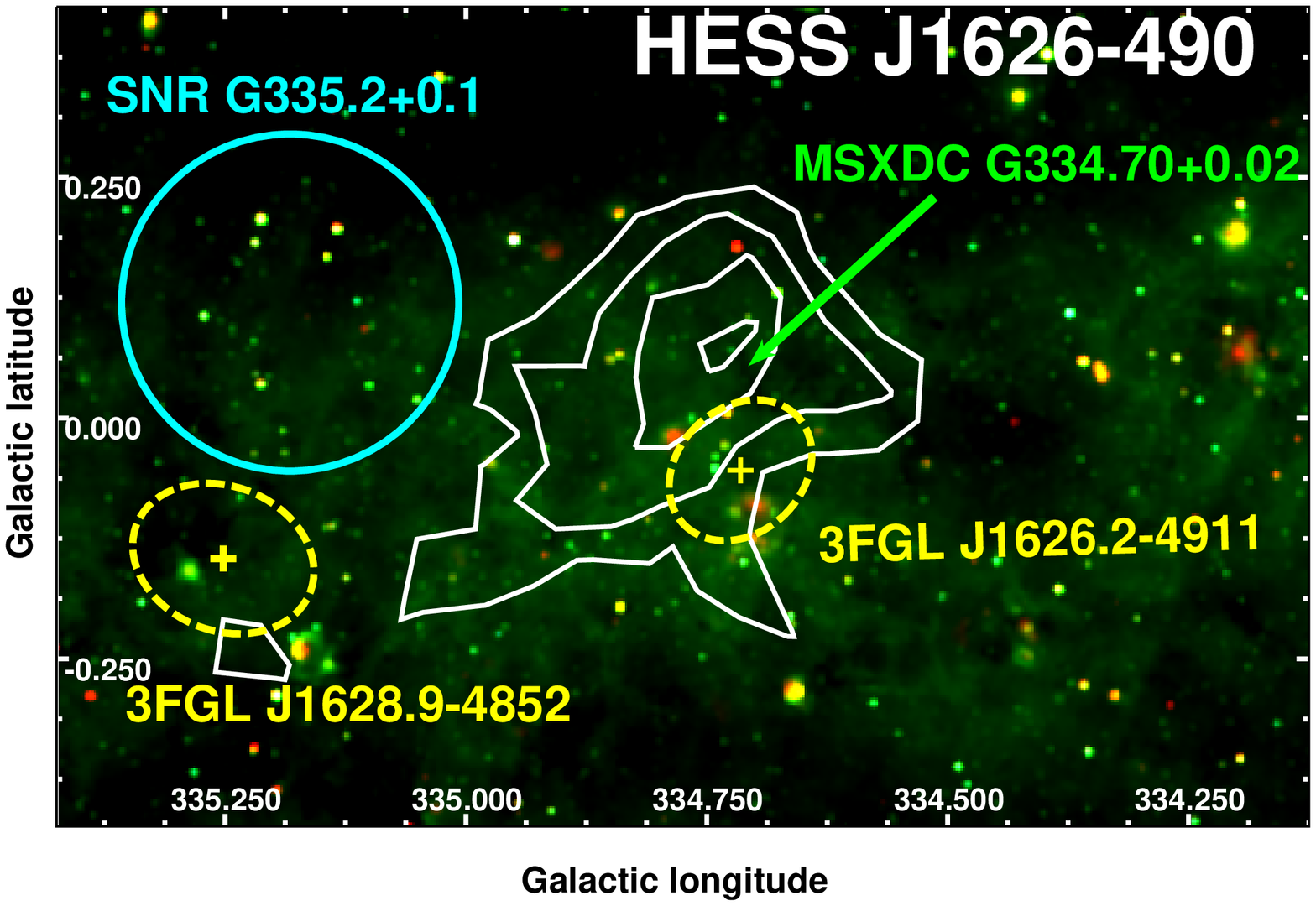}
\caption{\textit{Spitzer GLIMPSE/MIPSGAL} three-colour (RGB=24/8/4.5$\mu m$)M Jy sr$^{-1}$) image of the region towards HESS\,J1626$-$490. TeV emission is indicated by white contours, Fermi-LAT sources are indicted by dash, red ellipses, the SNR G335.2$+$0.1 is indicated by a cyan circle. Widespread 8$\mu\,m$ emission is seen, likely arising from poly-aromatic hydrocarbon (PAH) emission. Several small 24$\mu\,m$ emission can be seen as yellow sources, which traces heated dust. The IRDC MSXDC G334.70$+$0.02 is indicated towards the TeV emission.}
\label{J1626m490_IR_image} 
\end{figure*}

\subsection{HESS\,J1804$-$216}
\textbf{HESS\,J1804$-$216} is seen towards a number of interesting features including SNR G8.7$-$0.1 \citep{1986AJ.....92.1372O} and several H{\sc ii} regions observed as part of catalogues \citep{1997ApJ...488..224K,1996ApJ...472..173L,1982A&A...108..227W}. The high spin down power ($2.2\times10^{36}$\,erg/s) PSR J1803-2137 offers a plausible PWNe emission scenario for HESS\,J1804$-$216. However, the nature of the TeV source and its association with the X-ray sources found towards this object is still unclear \citep{2007ApJ...670..643K}. There is a 1720 MHz OH maser along the border of SNR G8.7$-$0.1. This maser signals a shock interaction between the SNR and dense clouds \citep{1998ApJ...508..690F} however, this maser is not coincident with the TeV emission as seen in Fig. \ref{J1804m216_image}. Several dense cloud clumps are seen in our study and are traced by the emission lines indicated in Table \ref{table:all_sources}. 

The dense gas seen in our study mainly surrounds the published extent of TeV emission. Thus, the morphology of the cool, dense gas supports a PWNe scenario for the TeV emission. However, there is infra-red poly-aromatic hydrocarbon (PAH) emission over the extent of the TeV emission, indicating that there is likely diffuse molecular material in this region not detected in our study. The lack of detected 12\,mm emission towards the 1720 OH maser suggests that there are undetected molecular clouds in our study, since molecular material with a density $\sim 10^5$\,cm$^{-3}$ is needed to produce 1720 MHz OH maser emission \citep{1999ApJ...525L.101W,1999ApJ...511..235L}. This additional molecular material as well as atomic material traced by H{\sc i} may provide enough target material for a hadronic origin of the TeV emission. Almost all NH$_3$\,(1,1) emitting clumps near HESS\,J1804$-$216 display H$_2$O maser emission (as seen in Fig. \ref{J1804m216_image}).

HESS\,J1804$-$216 is one of the brightest unidentified TeV sources and so may provide the opportunity for improved angular resolution studies and energy dependent morphology studies in TeV gamma-rays, potentially reducing confusion between counterparts at other wavelengths. Further studies of molecular gas, including tracers of warmer and more diffuse gas such as $^{12}$CO(1-0), CS (1-0) and H{\sc i} emission should be undertaken in this region to account for more of the gas mass than is implied by our studies. 

\begin{figure*}
\centering
\includegraphics[width=0.95\textwidth, angle=0]{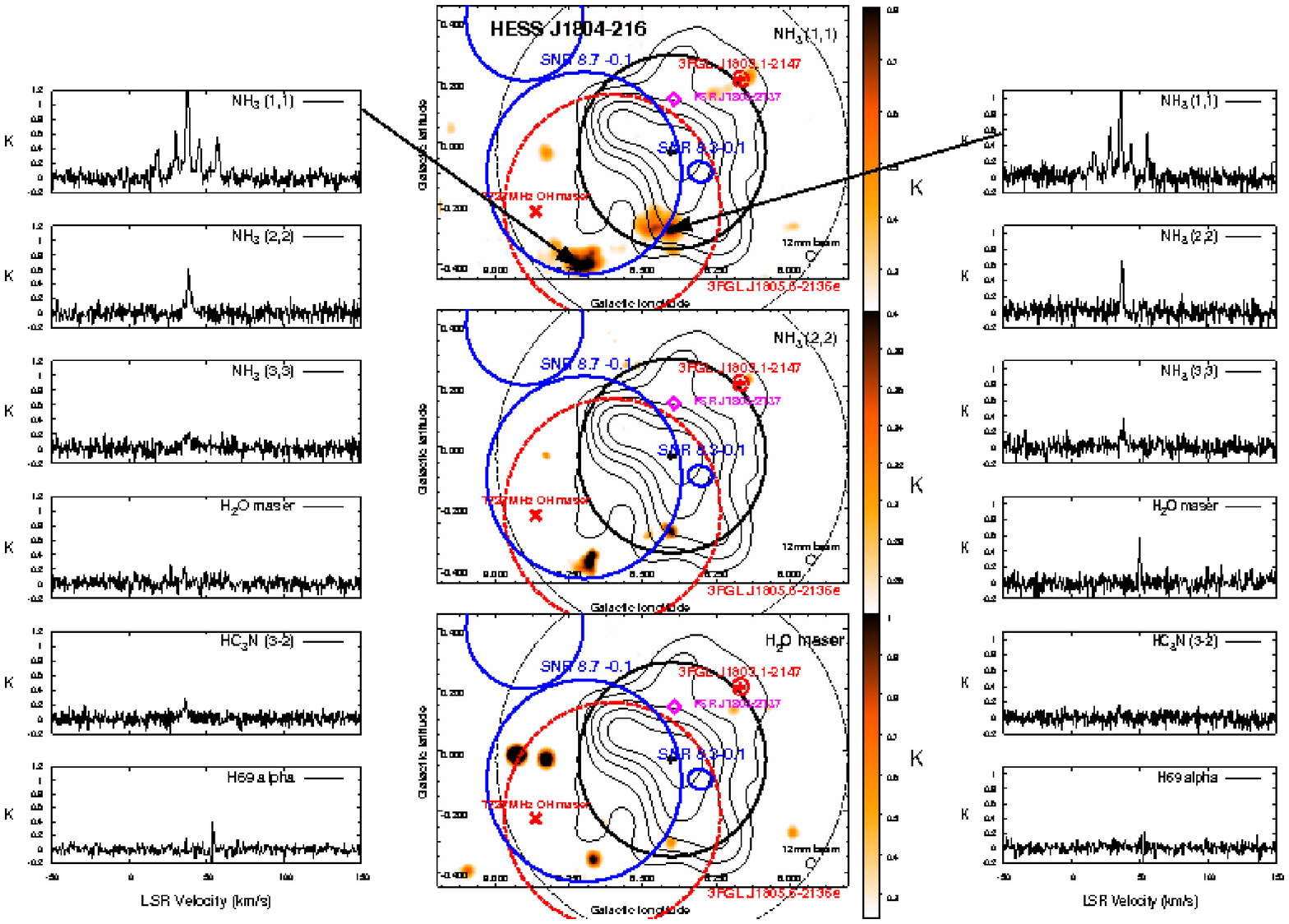}
\caption{12\,mm peak pixel maps of molecular line emission towards the TeV source \textbf{HESS\,J1804$-$216}. TeV emission is indicted by black contours, Fermi-LAT sources are indicted by dashed, red ellipses, the SNR G8.7$-$0.1 is indicated by a blue circle. The molecular emission detected in our study is limited to the edges of TeV emission. All molecular clumps detected show H$_2$O and/or H$69\alpha$ emission.}
\label{J1804m216_image}
\end{figure*}

\begin{figure*}
\includegraphics[width=0.8\textwidth]{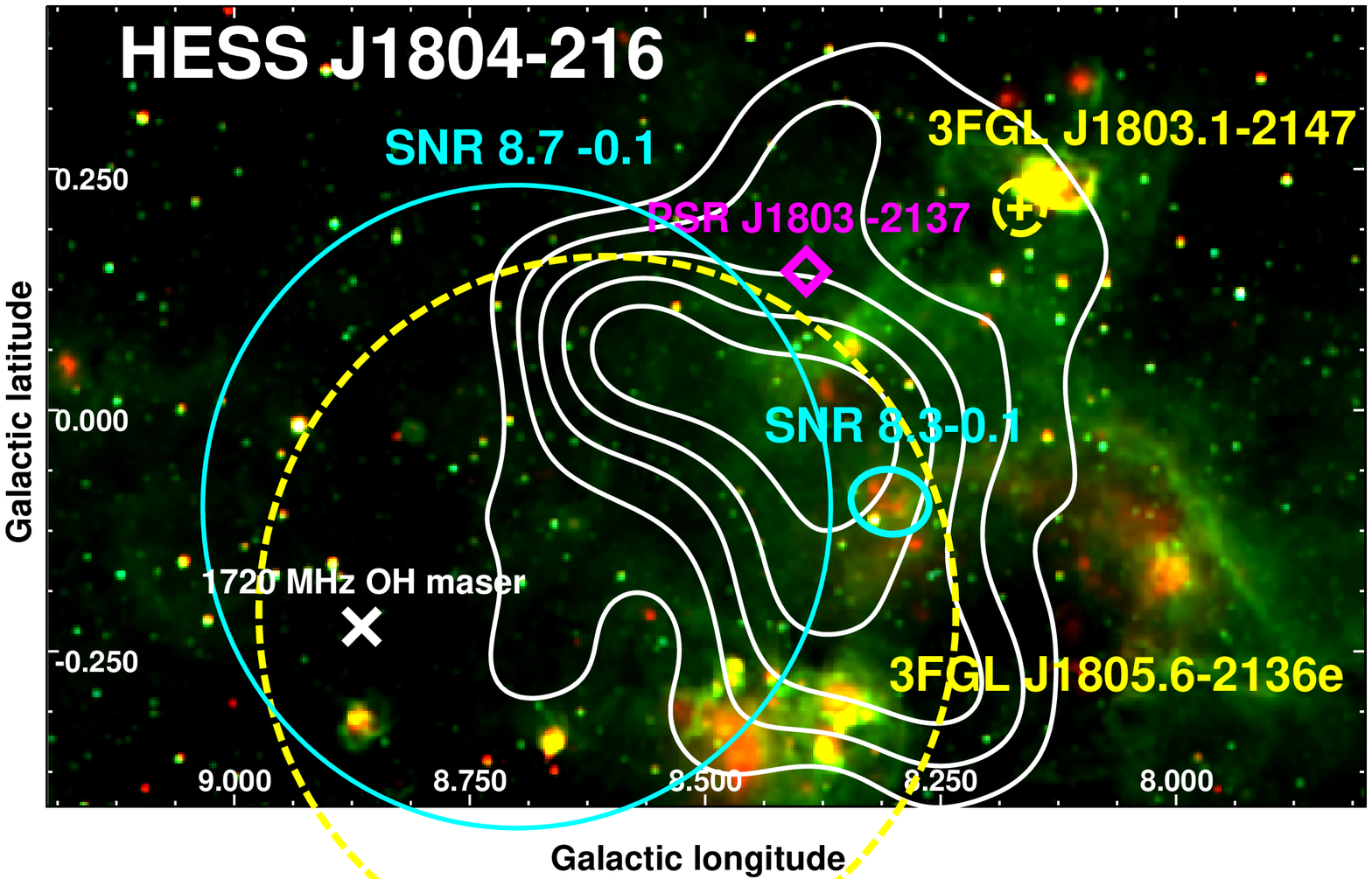}
\caption{\textit{Spitzer GLIMPSE/MIPSGAL} three-colour (RGB=24/8/4.5$\mu m$)M Jy sr$^{-1}$) image of the region towards HESS\,J1804$-$216. TeV emission is indicated by white contours, Fermi-LAT sources are indicted by dash, yellow ellipses, the SNR G8.7$-$0.1 is indicated by a solid, cyan circle and the pulsar PSR B1800$-$21 is indicated by a magenta diamond. Widespread 8$\mu\,m$ emission is seen, and numerous regions of heated dust (traced by 24 $\mu\,m$ emission) are seen, especially along the outside edge of the SNR, possibly indicating triggered star formation.}
\label{J1804_IR_image} 
\end{figure*}

\subsection{HESS\,J1729$-$345}
\textbf{HESS\,J1729$-$345} (aka HESS\,J1731$-$347 B) is offset from a TeV shell, HESS\,J1731$-$347 A \citep{2011A&A...531A..81H}, and lies near the direction of the H{\sc ii} region G353.381$-$0.114 catalogued in \citet{1987A&A...171..261C} at a distance of > 7.5 kpc which is not traced in this study. Previously, this H{\sc ii} region has been suggested as an indication for a molecular cloud complex, possible counterpart to HESS\,J1729$-$345. This molecular cloud complex is seen, traced by PAH emission in Fig. \ref{J1729m345_IR_image}. In this work a dense gas clump, traced with emission from the NH$_3$\,(1,1) transition is seen towards HESS\,J1729$-$345, at a velocity of $-16$\,km/s. This gas feature is seen in Fig. \ref{J1729_J1731_image} and also, as an IRDC in Fig. \ref{J1729m345_IR_image}. As this molecular cloud is seen as an IRDC, we can deduce that the NH$_3$ emission is foreground to the PAH emission which rules out the far kinematic distance solution of the NH$_3$ and tells us that the NH$_3$ cloud core is at 3.2 kpc. The kinematic distance of this gas feature is 3.8 kpc, within errors of a previous distance estimate of the SNR, based on it's assumed connection to the compact H{\sc ii} region G353.42$-$0.37, of 3.2 $\pm 0.8$kpc \citep{2008ApJ...679L..85T}. This is however foreground to a more recent SNR distance estimate of 5.2 - 6.2 kpc based on the assumption that the SNR is within the 3 kpc expanding arm \citep{2014ApJ...788...94F}. Further preliminary discussion about the distance of this source can be seen in \citet{Maxted:2015tsa} which details CS (1-0) emission.

If the molecular clump detected in this study was a counterpart to HESS\,J1729$-$345, the TeV emission could arise from $\pi^{0}$ decay resulting from p-p collisions between cosmic-rays accelerated in the SNR. The SNR is at a distance of 3.2 kpc \citep{2008ApJ...679L..85T}, as is the molecular clump towards HESS\,J1729$-$345 traced by NH$_3$\,(1,1) in our study. At this distance and with an angular separation between the centre of the SNR and HESS\,J1729-345 of $\sim0.5\degree$, the protons would have travelled approximately 30 parsecs. Using the method outlined in Section \ref{observations}, it would take $\sim$140 yrs for a $10^{14}$\,eV proton to diffuse from the SNR (HESS\,J1731$-$347) to HESS\,J1729$-$345 and $\sim$1400 yrs for a $10^{12}$\,eV proton. With a SNR age estimate of 27,000 yrs \citep{2008ApJ...679L..85T}, protons would have sufficient time to travel and interact with the molecular gas towards HESS\,J1729$-$345 and produce the TeV emission.

\begin{figure*}
\centering
\includegraphics[width=0.95\textwidth, angle=0]{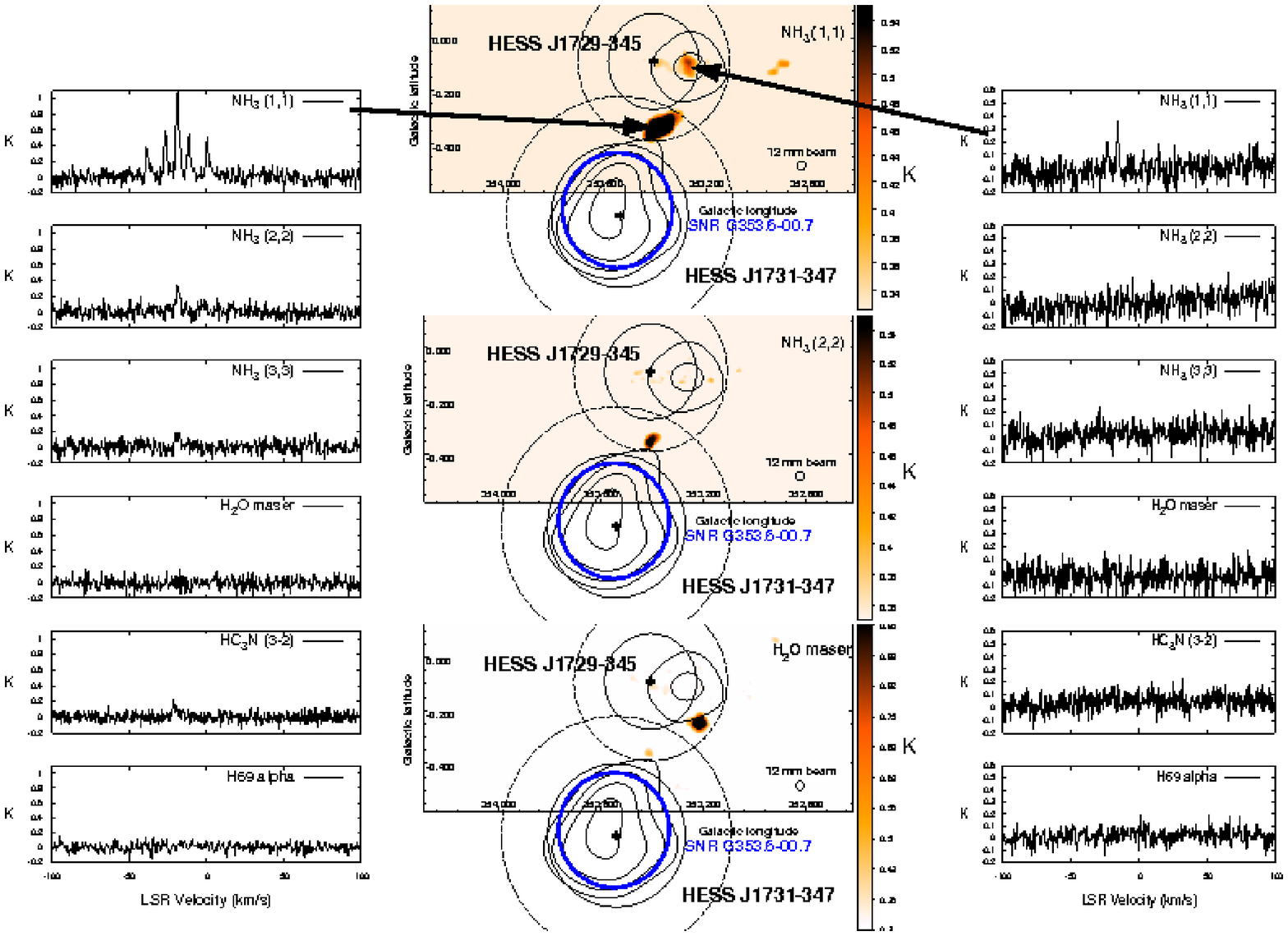}
\caption{12\,mm peak pixel maps of molecular line emission towards the TeV sources \textbf{HESS\,J1729$-$345} and HESS\,J1731$-$347. TeV emission is indicted by black contours, at the 4-7 $\sigma$ significance levels \citep[from][]{2011A&A...531A..81H}, and the defined 'towards' and 'adjacent' regions for HESS\,J1729$-$345 and HESS\,J1731$-$347 are represented by a black, thin, solid circle, and the area between this circle and a black, thin dashed circle respectively. The edge of the 12\,mm mapping region is indicated by the x-axis. Two molecular clumps can be seen, one towards the Galactic East from HESS\,J1729$-$345 at a velocity matching that identified for the SNR by \citet{2008ApJ...679L..85T}. The spectra for molecular transitions detected from this clump are seen on the RHS of the image.}
\label{J1729_J1731_image}
\end{figure*}

\begin{figure*}
\includegraphics[width=0.75\textwidth]{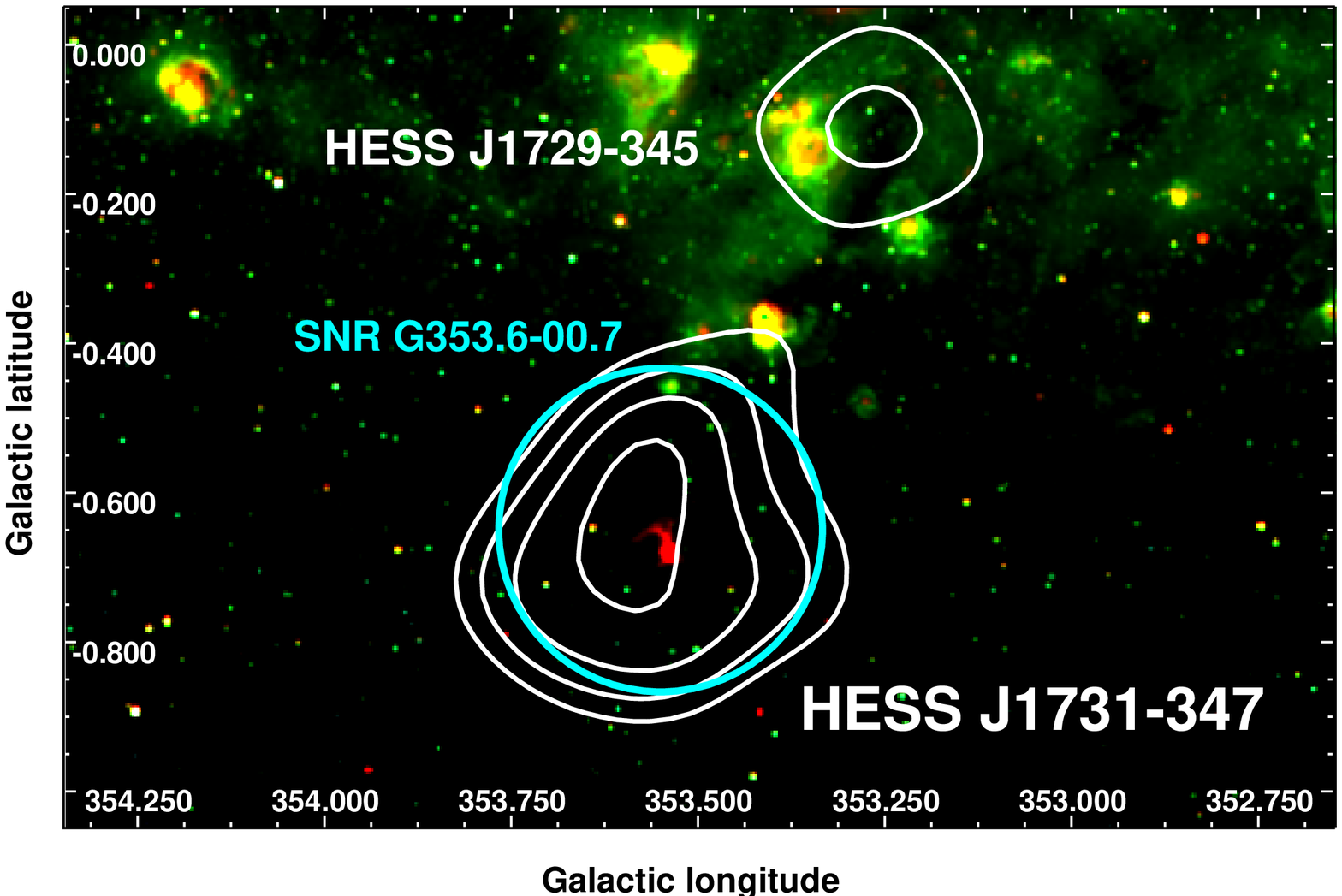}
\caption{\textit{Spitzer GLIMPSE/MIPSGAL} three-colour (RGB=24/8/4.5$\mu\,m$)M\,Jy\,sr$^{-1}$) image of the region towards HESS\,J1729$-$345 and HESS\,J1731$-$347. TeV emission is indicated by white contours. Widespread 8 $\mu\,m$ emission is seen towards HESS\,J1729$-$345 along with numerous regions of heated dust (traced by 24 $\mu\,m$ emission). This emission is assumed to be background to the molecular emission traced by our study which is towards part of the IRDC which can be seen extending from the SNR to HESS\,J1729$-$345.}
\label{J1729m345_IR_image} 
\end{figure*}

\section{Conclusions / Further Work}

The first large scale systematic study of dense (> 10$^4$\,cm$^{-3}$) gas towards Galactic TeV sources has been completed. HOPS equivalent coverage provides a good first look at distribution and dynamics of dense gas towards Galactic TeV sources. Knowledge about gas density profiles towards TeV sources allows for more robust studies of cosmic ray diffusion. Preliminary results suggest 12\,mm NH$_3$ inversion transitions, used to estimate NH$_3$ ortho-para ratios, could be used to search for regions of dense gas with previous shock activity. With this first look, we have found dense gas counterparts to unidentified regions of TeV emission including HESS\,J1745$-$303B and HESS\,J1745$-$303C as well as HESS\,J1848$-$018 and HESS\,J1626$-$490 which indicate it is likely there is a hadronic component to these sources.

Further observations of TeV sources showing dense gas overlap traced by the NH$_3$\,(1,1) emission seen in the H$_2$O Galactic Plane Survey would be beneficial at millimetre wavelengths. Further 12\,mm observations would provide more sensitivity to dense gas clumps traced by NH$_3$ emission. Observations at 7\,mm to trace molecular transitions such as SiO (1-0) which traces shocked gas and CS (1-0) which traces the denser features of gas clouds will provide better understanding of the mechanisms producing the relativistic particles producing the TeV gamma-rays as well as understanding the kinematics of the densest parts of gas towards these TeV gamma-ray sources. Further 12\,mm observations which include the NH$_3$\,(4,4) and (5,5) transitions would be beneficial to estimate the NH$_3$ OPR in sources which we have found an anomalous NH$_3$\,(3,3)-to-(1,1) brightness temperature ratio.

In addition, studies of particle transport within individual TeV sources should include not only the dense gas described here, but observations of moderately dense molecular gas, e.g. that traced by CO(1-0), and atomic gas, traced by HI, in order to understand density profiles of molecular clouds towards and adjacent to the TeV emission. The Mopra CO Survey \citep{2013PASA...30...44B} provides a large scale study for more diffuse gas towards Galactic TeV sources. With an angular resolution of 33 arcsec, and the inclusion of several isotopologues (including  which help to trace a range of molecular gas densities, the Mopra CO Survey will provide the most accurate large scale look at molecular gas in the Milky Way for use with current and future gamma-ray data sets. The Mopra CO and HOPS angular resolution (33 arcsec and 1 arcmin respectively) and the future CTA angular resolution (2-3 arcmin between 1-10 TeV) are comparable and have the ability to resolve molecular cloud cores. In addition, all of these observations will cover a significant portion of the Galactic plane allowing for improved morphological comparison studies, and robust statistical studies of molecular gas and TeV gamma-rays. These higher resolution gas surveys will also provide a necessary picture of the distribution of Galactic gas to be used for new gamma-ray emission templates needed for higher angular resolution gamma-ray observations. HI gas is also an important contribution to the target gas mass for cosmic rays \citep[e.g.][]{2012ApJ...746...82F}. Quantifying the contribution from HI has been improved with recent studies of optically thick HI in the ISM \citep[e.g.][]{2015ApJ...798....6F}. The contribution of dark gas (both molecular and atomic) may also be important \citep[e.g.][]{2015ApJ...811...13B}. 

Currently, a robust statistical study of the overlap between Galactic TeV emission and star forming regions is not possible due to confusion, sensitivity, and uneven coverage. Recent observations of the Large Magellenic Cloud \citep{2015Sci...347..406H} have shown that superbubbles are capable of producing TeV emission, and it is suspected that we will find similar Galactic environments. As the angular resolution of TeV instruments improves, specifically with CTA, to match that of molecular gas studies (such as HOPS and MopraCO) and the sensitivity of molecular gas observations improves also, a much better understanding of the emission mechanisms and high energy particle transport in Galactic star forming regions will evolve.

\section*{Acknowledgements}
This work was supported by an Australian Research Council grant (DP1096533). The Mopra Telescope is part of the Australia Telescope and, at the time of these observations, was funded by the Commonwealth of Australia for operation as a National Facility managed by CSIRO. The University of New South Wales Mopra Spectrometer Digital Filter Bank used for these Mopra observations was provided with support from the Australian Research Council, together with the University of New South Wales, University of Sydney, Monash University and the CSIRO. This research has made use of NASA's Astrophysics Data System Bibliographic Services, the SIMBAD database, operated at CDS, Strasbourg, France \citep{2000A&AS..143....9W} and the ATNF Pulsar Catalogue \citep{2005AJ....129.1993M} which can be found at http://www.atnf.csiro.au/people/pulsar/psrcat/.




\bibliographystyle{mnras}
\bibliography{tev_dense_gas_overlap_mnras} 

\bsp	
\label{lastpage}
\end{document}